\documentclass[preprint2]{aastex}
\usepackage{latexsym, amsmath, amssymb,amsthm,amsopn,amsfonts, graphicx, epstopdf}
\usepackage{natbib}

\newcommand{\figureformat}{pdf}

\slugcomment{submitted to ApJ supplement}

\shorttitle{MADmap}
\shortauthors{Cantalupo et al.}

\begin{document}

\title{\vspace*{-0.75in}MADmap:  A Massively Parallel Maximum-Likelihood\\ Cosmic Microwave Background Map-Maker}

\author{\vspace*{-0.2in}C. M. Cantalupo}
\affil{Computational Cosmology Center, Lawrence Berkeley National Laboratory, \\ Berkeley, CA 94720, USA}
\email{cmcantalupo@lbl.gov}
\author{\vspace*{-0.2in}J. D. Borrill}
\affil{Computational Cosmology Center, Lawrence Berkeley National Laboratory, \\ Berkeley, CA 94720, USA}
\email{jdborrill@lbl.gov}
\author{\vspace*{-0.1in}A. H. Jaffe}
\affil{Department of Physics, Blackett Laboratory, Imperial College, \\ London SW7 2AZ, Great Britain}
\email{a.jaffe@imperial.ac.uk}
\author{\vspace*{-0.2in}T. S. Kisner}
\affil{Computational Cosmology Center, Lawrence Berkeley National Laboratory, \\ Berkeley, CA 94720, USA}
\email{tskinser@lbl.gov}
\author{\vspace*{-0.2in}R. Stompor}
\affil{CNRS, Laboratoire AstroParticule et Cosmologie, Universit{\'e} Paris-7 `Denis Diderot', \\75205 Paris Cedex 13, France}
\email{radek@apc.univ-paris-diderot.fr}

\begin{abstract}
MADmap is a software application used to produce maximum-likelihood
images of the sky from time-ordered data which include correlated
noise, such as those gathered by Cosmic Microwave Background (CMB)
experiments. It works efficiently on platforms ranging from small
workstations to the most massively parallel supercomputers. Map-making
is a critical step in the analysis of all CMB data sets, and the
maximum-likelihood approach is the most accurate and widely applicable
algorithm; however, it is a computationally challenging task. This
challenge will only increase with the next generation of ground-based,
balloon-borne and satellite CMB polarization experiments. The
faintness of the B-mode signal that these experiments seek to measure
requires them to gather enormous data sets. MADmap is already being
run on up to $O(10^{11})$ time samples, $O(10^8)$ pixels and $O(10^4)$
cores, with ongoing work to scale to the next generation of data sets
and supercomputers.  We describe MADmap's algorithm based around a
preconditioned conjugate gradient solver, fast Fourier transforms and
sparse matrix operations. We highlight MADmap's ability to address
problems typically encountered in the analysis of realistic CMB data
sets and describe its application to simulations of the Planck and
EBEX experiments. The massively parallel and distributed
implementation is detailed and scaling complexities are given for the
resources required. MADmap is capable of analysing the largest data
sets now being collected on computing resources currently available,
and we argue that, given Moore's Law, MADmap will be capable of
reducing the most massive projected data sets.
\end{abstract}

\keywords{Cosmic Microwave Background, data analysis, map making, maximum likelihood}

\section{Introduction}
The Cosmic Microwave Background (CMB) is the left-over radiation from
the Big Bang. This radiation is comprised of primordial photons
last scattered when the first neutral atoms formed
some 400,000 years after the Big Bang.  The subsequent expansion of
the Universe redshifts the spectrum of these photons, originally
distributed as a 3000K black body, to appear in contemporary
observations as a 3K black body.  Encoded within the image of the
thermal fluctuations in the CMB are not only the signatures of the
basic parameters of cosmology but also, using the Big Bang as the
ultimate particle accelerator, insights into fundamental physics at
the very highest energies \citep{bib:dodelson03}.

Most CMB experiments simply scan across the sky, sampling its
temperature (and now polarization) at one or more microwave
frequencies at some fixed rate. Once these data have been cleaned and
calibrated, the first step in their scientific analysis
involves producing pixelized maps of the sky and at least an estimate
of the pixel noise properties. Subsequent steps include separating
these sky maps into distinct CMB and foreground component maps,
estimating the power spectra of the CMB from its maps, and estimating
cosmological parameters from these power spectra. Since these
subsequent steps depend on the quality of the original maps, it is
important to generate the most accurate and well-characterised maps we
can, in particular maximum-likelihood, minimum variance
maps. However, the faintness of the CMB fluctuations on the sky drives
us to gather enormous data sets which must be reduced coherently if we
are to preserve their scientific content, resulting in a very
significant computational challenge.

The computational costs of the dominant CMB analysis steps (in
floating-point operations, memory, communication, and input/output)
are set by the numbers of time samples ${\rm n_t}$ and sky pixels ${\rm n_z}$. As the goals of CMB
experiments have evolved, both of these numbers have grown.  The
growth in the number of samples is driven by the quest for fainter polarized and high
multipole signals, and the number of pixels is driven by the sky-coverage and
resolution required for measuring low and high multipole signals
respectively. Table~(\ref{tab:dvol}) shows this evolution for a
representative selection of suborbital and all anticipated satellite
CMB missions.

\begin{table*}[htbp]
 \centering
 \begin{tabular}{|c|c|c|c|c|c|}
 \hline
 Date & Experiment & Description & Samples (${\rm n_t}$) & Pixels (${\rm n_z}$)\\
 \hline
 1990-93 & COBE & All-sky, low-res, T & $8 \times 10^{8}$ & $3 \times 10^3$ \\
 1998 & BOOMERanG & Cut-sky, mid-res, T & $9 \times 10^{8}$ & $3 \times 10^5$ \\
 2001-10 & WMAP  & All-sky, mid-res, TE & $2 \times 10^{11}$  & $6 \times 10^6$ \\
 2009-11 & Planck & All-sky, high-res, TE & $3 \times 10^{11}$ & $1 \times 10^8$ \\
 2010 & EBEX & Cut-sky, high-res, TEB & $3 \times 10^{11}$ &  $6 \times 10^5$ \\
 2010-12 & PolarBeaR & Cut-sky, high-res, TEB & $3 \times 10^{13}$ & $1 \times 10^7$ \\
 2010-13 & QUIET-II & Cut-sky, high-res, TEB & $1 \times 10^{14}$ & $7 \times 10^5$ \\
 2020+ & CMBpol & All-sky, high-res, TEB & $1 \times 10^{15}$ & $9 \times 10^8$ \\
 \hline
 \end{tabular}
 \caption{The evolution of sample and pixel counts over time for a
   representative selection of suborbital and all anticipated
   satellite missions. Details of the proposed CMBpol satellite are
   from the EPIC Intermediate Mission concept study. To allow 
   meaningful comparison, the {\it effective} number of samples and pixels 
   for each experiment are given, which may differ form those 
   quoted by the respective experiment teams.  
   Specifically, sample
   counts are defined as the sum over all of an experiment's
   time-ordered data streams of the stream's sampling rate times the
   duration of its observation (where data streams may include various
   combinations of detector streams), while pixel counts are the sum
   over all observing frequencies of the fraction of the sky observed
   divided by the fiducial beam size, assuming a circular beam with
   diameter given by the assumed full-width at half-maximum, at that 
   frequency.}
   \label{tab:dvol}
\end{table*}

Since brute force dense matrix maximum-likelihood algorithms scale as $O({\rm n_z}^3)$
\citep{bib:borrill99} they have become computationally intractable, and we have
had to adopt alternative approximate algorithms that are dominated by operations on the time-ordered data that are
linear and log-linear in ${\rm n_t}$. Over the next
15 years we can expect CMB time-ordered data volumes to grow by 3
orders of magnitude; coincidentally this exactly matches the projected
growth in computing power over the same period assuming a continuation
of Moore's Law. Since CMB data analysis is already pushing the limits
of current state-of-the-art HPC systems, this implies that our
algorithms and implementations will have to continue scaling on the
leading edge of HPC technology for the next 10 Moore-doublings if we
are to be able fully to support first the design and deployment of
these missions and then the scientific exploitation of the data sets
they gather.

The next section gives an overview of the computational
context for the MADmap software: the libraries it depends on, the
supporting applications, and hardware platforms targeted.
Section~\ref{sec:formalism} describes the formalism which is applied
to the data by MADmap including the statistical derivation and
mathematical framework.  MADmap is designed around the method of
preconditioned conjugate gradients, and Section~\ref{sec:algorithm}
describes this algorithm. Section~\ref{sec:implementation} details MADmap's implementation, including
subsections on data distribution, noise weighting, pointing data compression, pixel indexing, and the functional complexity of the
communication, memory, CPU and disk requirements of MADmap.  Section~\ref{sec:using} explores the versatility of the MADmap data
model, and describes a variety of real world problems that MADmap has
been used to solve.  Section~\ref{sec:examples} describes a set of MADmap runs on
simulated Planck and EBEX data and explains in detail the performance
characteristics of these runs which span a range of data sizes in both
the time and pixel domain and a variety of processor counts.  The
paper finishes with a comparison with other software, a discussion of
future work, and concluding remarks.

We will use the following notation in this paper:
variables in unitalicized roman font are scalars (e.g., ${\rm n_t}$ and ${\rm n_i}$),  
time domain vectors are represented by italic Greek letters (e.g., $\gamma$ and $\nu$),  
pixel domain vectors are represented by italic Roman letters (e.g., $y$ and $z$), and  
matrices are capitalized (e.g., $A$ and $B$).  

\section{MADmap} \label{sec:madmap}

Under the assumption of piecewise stationary Gaussian noise (defined more precisely below) with
known spectral properties MADmap produces maximum-likelihood maps
to user-specified precision, and in particular does so for the very
largest real and simulated CMB data sets extant and on the very
largest of today's supercomputers.  For example, MADmap can enable a wall-clock
time to solution of tens of minutes for Planck-like data volumes running on a significant fraction of 
the 40,000-core Cray XT4 at the US Department of Energy's National
Energy Research Scientific Computing Center. Supercomputers will be a 
crucial resource in the analysis of forthcoming CMB data sets and 
MADmap's ability to scale to use large computing resources will enable 
science that would be otherwise inaccessible~\citep{bib:bock06}.  The maps' noise
properties can be calculated by other tools in the MADCAP software
suite. MADpre, an application distributed with MADmap, 
generates the auto-correlation matrix assuming time domain white noise
(which can be used as a precondioner by MADmap), while MADping
calculates the full dense noise covariance matrix by explicit
inversion, although this is impractical for more than $O(10^5)$
pixels.

These software components take advantage of a set of libraries to
facilitate simulation and input/output operations.  M3 is a data
abstraction library that uses a plug-in architecture to enable analysis applications to ingest data
from a variety of experiments, which invariably adopt different data formats and distributions. Crucially, M3 can also simulate simple signals and complex noise on the fly, which reduces the potentially overwhelming I/O and disk cost of the traditionally separate simulation and analysis steps.
Similarly, the Generalized Compressed Pointing (GCP) library,
described in more detail below, has been devised to calculate the pointings of 
individual detector samples on demand at run time from compressed
pointing information (e.g., the sparse-sampled pointing of a reference frame, such as the focal-plane
bore-sight).

MADmap was originally designed to map CMB data in a manner
independent of any individual CMB experiment, and has been applied to
real and simulated data from a number of missions including the
historic BOOMERanG~\citep{bib:debernardis00} and MAXIMA~\citep{bib:hanany00} 
experiments, and the upcoming 
EBEX\footnote{EBEX: {\tt http://groups.physics.umn.edu/cosmology/ebex/}}~\citep{bib:oxley04} balloon and 
Planck\footnote{Planck: {\tt  http://www.esa.int/esaSC/120398\_index\_0\_m.html}} 
satellite missions. However the problem it solves
appears in any situation where a signal described by a sparse linear
model is to be derived from data containing correlated Gaussian noise
that is stationary over known intervals, and it is already being used
in other domains, for example by the Herschel mission~\citep{bib:waskett07}.

We also note that map-making-like operations are commonly used in
other stages of the data analysis, including power spectrum 
estimation via sampling algorithms~\citep{bib:jewell04, bib:wandelt04} and
component separation via parametric technique~\citep{bib:stompor09}, 
and MADmap can be employed in a straightforward manner in all those contexts.

\section{Formalism} \label{sec:formalism}

The central assumption of our map-making is that the time-ordered data measured by each detector,
$\gamma$, can be written
\begin{eqnarray}
\label{eDSN}
\gamma = \nu + \zeta \\
\zeta = A z
\end{eqnarray}
for temporal noise $\nu$, pixelized
signal $z$, and a sparse pointing matrix $A$ encoding the weights
with which each pixel is observed at each time.  In the
simplest case (a total power CMB temperature with uniform symmetric
beams in the limit of  small pixels compared to the beam size) the pointing matrix contains a single unit weight per
observation; at time sample $t$ with the detector pointed at sky-coordinates
$(\theta_t, \psi_t)$
\begin{equation}
A_{t,p} = \left\{ \begin{array}{ll}
	        1, & {\rm if} \;\;\; (\theta_t, \psi_t) \in {\rm pixel\ p}; \\
                0, & {\rm otherwise.}
	                 \end{array}
                \right.
\end{equation}
For ideal polarization experiments, this single weight is then
replaced by one for each of the Stokes parameters in the
observed pixel:
\begin{equation}
\zeta_t = \frac12\left[ i_p + q_p\sin(2 \alpha_t) + u_p\cos(2\alpha_t)\right]
\end{equation}
where $\alpha$ is a time ordered vector of the observed polarization angle, and we consider the signal as a vector $ z=[i, q, u]$.
More complex beams may be incorporated by
appropriately weighting multiple pixels in each observation, and
parasitic signals that are fixed in any other basis (for example,
MAXIMA's chop-synchronous signal~\citep{bib:stompor02}) can be 
simultaneously solved for by extending the pixel basis appropriately
(see also Section~\ref{sec:using}). 
Ultimately the only requirement on $A$ is that it be 
full rank, which is to say that there are at least as many linearly independent observations
as there are signals that are being solved for.  
In the broadest sense the pointing matrix is
a projection from the signal basis to the time basis that 
defines the linear relationship between the signal to be solved for and 
the recorded detector time stream.  

Representing now the noise as,
\begin{equation}
\nu = \gamma - A z
\end{equation}
and making use of its Gaussian properties we can write the likelihood function of 
a pixelized sky map $z$ given our data $\gamma$ as  
\begin{equation}
{\cal L}(z|\gamma) \propto \exp \left\{ -\frac{1}{2} \left(\nu^T  N^{-1} \nu + {\rm Tr} \left[ \ln N \right] \right) \right\}\;,
\end{equation}
where we have used the matrix identity 
\begin{equation}
\mathrm{Tr} \ln H =\ln \mathrm{det} H
\end{equation}
to move the usual Gaussian prefactor into the exponential.
Maximizing this over all {\em a priori} equally likely signals
$z$, gives the maximum-likelihood map and its noise correlation matrix
\begin{eqnarray}
\label{eMLM}
\hat{z} & = & M A^T N^{-1} \gamma \nonumber \\
M & = & \left( A^T N^{-1} A \right)^{-1}
\end{eqnarray}
which can also be derived as the minimum variance or generalized
least squares solution~\citep{bib:degasperis05}. 
In addition, the map ${\hat z}$ is a sufficient statistic for the likelihood 
functions as written. That is, the Gaussian likelihood of a sky map $z$ only depends on the estimate ${\hat z}$ and no other function of the data: 
\begin{equation}
{\cal L}(z|\hat{z})\propto \exp \left\{ -\frac{1}{2} \left[ (z - {\hat z})^T  M^{-1} (z-{\hat z}) + {\rm Tr} (\ln M )\right]\right\}
\end{equation}
where the map noise correlation matrix is $M$.  

These equations reduce to simple averaging into pixels in the white noise limit, $N_{t,t'}=\sigma^2\delta_{t,t'}$.
We note that the $O({\rm n_z}^3)$ scaling of the brute force explicit dense matrix calculation
arises from the inversion of the inverse pixel noise matrix, although it is important to note that the computational cost of building the matrix 
before inversion, $O({\rm n_t} {\rm n_c})$, may dominate $O({\rm n_z}^3)$ for some experiments.

Our solution $\hat{z}$ is only the maximum-likelihood estimate of $z$
only if $N$ is the true noise correlation matrix, and since the noise
properties of CMB data have to be derived from the data there will
inevitably be uncertainties. However, provided $N$ is positive
definite, then $\hat{z}$ provides an unbiased, though potentially
not optimal, estimate of $z$.

\section{Algorithm} \label{sec:algorithm}

Brute force explicit dense matrix computation of  Equation~(\ref{eMLM}), requiring the construction
and inversion of the full dense pixel-pixel noise correlation matrix,
is computationally impractical for current and future CMB data sets
with millions to hundreds of millions of pixels. However, we observe
that it can be re-written in standard linear form~\citep{bib:oh99},
\begin{equation}
\label{eq:pcgSystem}
H x = b
\end{equation}
where $H \equiv A^T N^{-1} A$,  $b \equiv A^T N^{-1} d$ and $x \equiv \hat{z}$. 

Since $N^{-1}$ is positive definite (being a correlation matrix) and
$A$ is full rank (by assumption, but see the discussion later in this section), $H$ is also guaranteed to be
positive definite. Now provided we can operate on any time-domain
vector with $N^{-1}$ (weighting) and $A^T$ (pointing), and on any
pixel-domain vector with $A$ (unpointing), we can solve for $x$ using
preconditioned conjugate gradient (PCG) - a widely-used technique for
solving positive definite linear systems that provides the fastest
time to solution for many problems (especially sparse systems),
\citep{bib:barrett94, bib:shewchuk94}).

The method of conjugate gradients is an optimization algorithm similar
to steepest decent except that search directions are chosen to be
orthogonal to each other.  This optimization is applied to minimize
the quadratic form:
\begin{equation}
g(x) = \frac{1}{2} x^{T} H x - b^{T}x + c
\end{equation}  
which has the following derivative with respect to $x$:
\begin{equation}
g'(x) = H x - b.
\end{equation}
When $H$ is positive definite $g$ is a convex quadratic, so that when
the gradient of $g$ is zero we have both minimized $g$ and solved the
original linear system of Equation~(\ref{eq:pcgSystem}). Embedding the
solution of a positive definite linear system into the optimization of
a convex quadratic may at first seem just to complicate matters, but
operationally it buys us a lot if we can operate with $H$ on a vector
quickly and with minimal memory overhead. This is an iterative
technique, and each iteration involves the operation of $H$ on a
vector, which is the dominant computational cost of the
algorithm. Since the number of iterations required to converge to the
solution is proportional to the condition number of $H$, having an
approximate inverse of $H$ that can multiply a vector quickly can
greatly reduce the number of iterations required by making the
effective matrix in the preconditioned system approximate the identity
matrix. 

The time streams collected by CMB telescopes generally contain large
low frequency correlations in the noise which is often referred to as 
``1/f-noise''. The noise power spectrum [the Fourier transform of the noise correlation function $n(|t-t'|)$] can be modelled as
\begin{equation}
p(f) = \sigma^2 \left[1+(f_k/f)^\alpha \right]
\end{equation}
with a white noise level $\sigma$, knee-frequency $f_k$, and typically
power law $1 \leq \alpha \leq 2$. Having power
inversely proportional to frequency clearly blows up for the constant
mode ($f = 0$)\footnote{We note that for real detectors the power at zero frequency
is clearly finite, though still high enough to lead to a numerical near degeneracy.}.  
This (near) degeneracy produces a null space to the
constant mode in the inverse pixel pixel noise correlation matrix, and
therefore the inverse problem is ill-posed (equivalently, the $H$ matrix is only positive \emph{semi}-definite).  Applying the conjugate
gradient technique to such ill-posed problems has been discussed in the
literature \citep{bib:brakhage96}, and used in many
fields. In the specific example discussed
here, the total offset of the estimated map is arbitrary.  In a more
general case such degeneracies may result from the presence of (or
removal of) a systematic effect in the time-ordered data, and need to
be considered case by case.  In general when applying conjugate 
gradient to semi-definite linear systems the iterates will begin by converging 
toward the optimal solution until a point after which subsequent iterations begin to diverge from the 
solution.  Choosing when to stop iterating is the key to dealing with ill-posed problems,
and this is discussed in more detail in \citet{bib:brakhage96}.  

Observations of CMB polarization can also introduce degeneracies if a
particular pixel is not observed with sufficiently many different
orientations to separate the 3 Stokes parameters. This situation can
be resolved by identifying such pixels ahead of time (for example by
examining the conditioning of the $3\times3$ blocks, or an appropriate
generalization of those if different pixel resolutions are adopted for different
Stokes parameters, in the $A^T A$ matrix)
and excising poorly-conditioned pixels.

Removing pixels from the map and therefore their respective samples
from the time stream affects the structure of the time-domain noise
correlation matrix, making it non-Toeplitz, and thus difficult to
handle efficiently, whenever the noise is correlated. This problem can
be overcome by the so-called gap filling
procedure~\cite{bib:stompor02}, which within the MADCAP software suite
can be facilitated by the MADmask and MADnes applications. These tools
replace time ordered data samples that occur while an excised pixel is
observed by simulated noise that is consistent with the power spectrum
associated with each excised sample and the correlations inferred by
the noise in surrounding samples.  In this way the continuity and
stationarity of the time-domain noise is preserved, together with the
Toeplitz character of its respective noise correlation matrix. The
rows of the pointing matrix that correspond to observations of excised
pixels are set to zero, and this models the simulated noise as free of
signal.

\section{Implementation} \label{sec:implementation}

The core of MADmap is a massively parallel implementation of a
preconditioned conjugate gradient (PCG) solver. 
Each PCG iteration moves in a direction in the solution space that is
orthogonal to all previous steps.  In the absence of numerical 
error and degeneracies in the system the exact solution is guaranteed
after after ${\rm n_z}$ steps because the number of possible directions has been
exhausted. In practice however the
calculation is terminated either after a fixed number of iterations
${\rm n_i}$, or after achieving a given accuracy as measured by the relative
residual $\epsilon$ (in the absence of user-specified values MADmap
defaults to ${\rm n_i} = 50$, $\epsilon = 10^{-6}$). On termination MADmap
outputs the vector which produced the lowest relative residual thus
far, which is not necessarily the last iteration of the PCG as the
relative residual is not guaranteed to decrease monotonically with
iterations.  By algorithmic design the relative error does not increase
with iteration. 

It is possible to precalculate a sparse preconditioner and have MADmap read
this preconditioner in from disk at run time; for example, the MADpre code will precompute and 
store $(A^T W^{-1} A)^{-1}$, where $W^{-1}$ is simply $N^{-1}$ with all off-diagonal elements set to zero. 
This is the noise correlation matrix if white noise is assumed in the 
time domain. If no preconditioner file 
is specified then by default MADmap calculates the point Jacobi preconditioner 
which is the diagonal matrix composed of the inverse of the diagonal elements of $A^T A$.

\begin{figure*}[htbp]
\begin{center}
\includegraphics[scale=0.6]{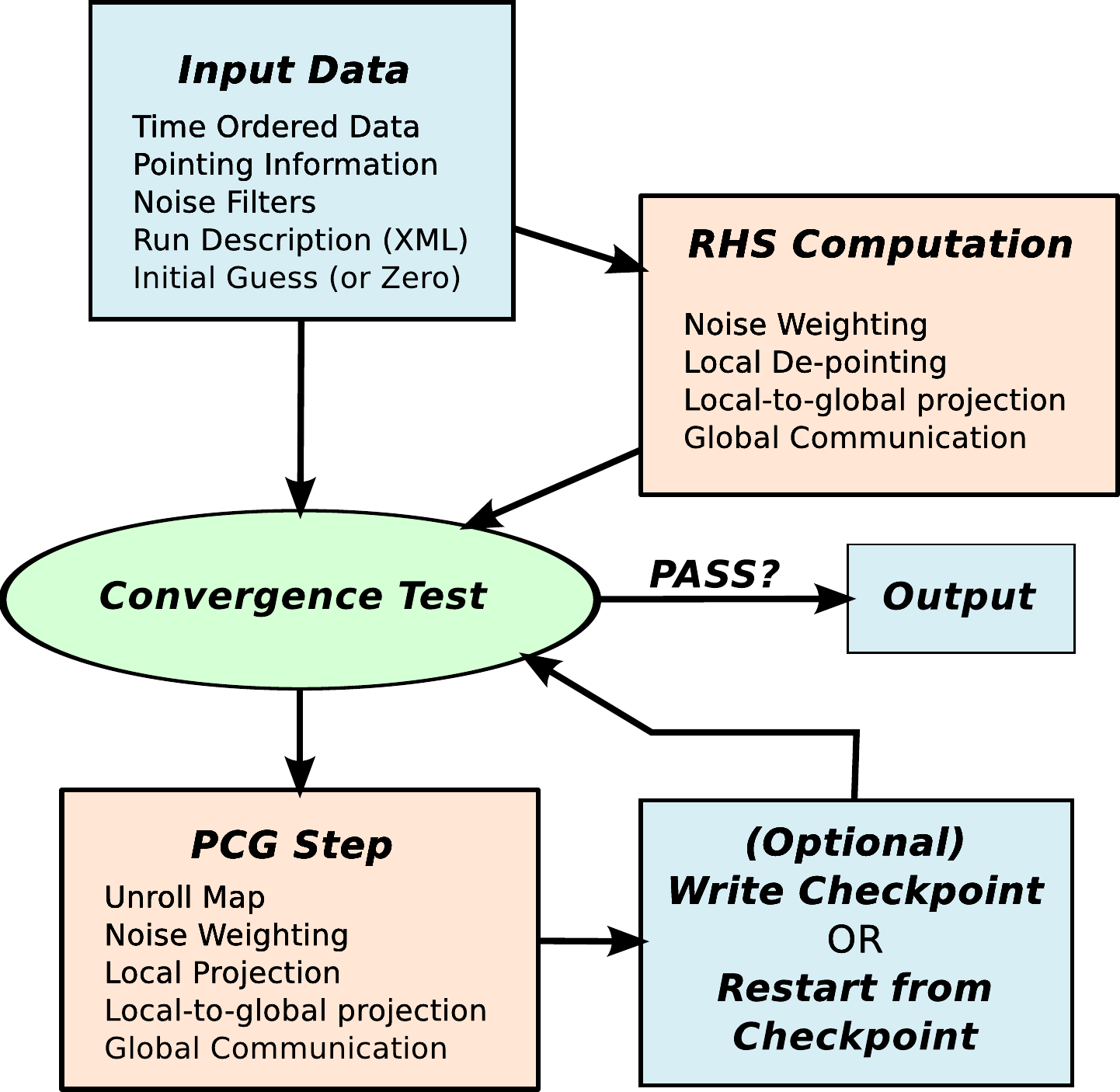}
\caption{A graphical overview of MADmap.  Blue boxes represent data products and pink boxes represent computational procedures, where RHS is the ``right hand side'' calculation.
\label{mapflow}
}
\end{center}
\end{figure*}

To minimize the impact of system failure on a calculation in progress,
MADmap checkpoints periodically and can restart from any of these. By
default MADmap checkpoints every 20 PCG iterations, although this can
be altered by setting an environment variable\footnote{The checkpoint 
frequency environment variable: {\tt MM\_CHECKPOINT\_FREQUENCY}}.  
Each checkpoint comprises five map
vectors and six scalars, with the distributed maps being gathered onto
the root processor before being written to disk. At a checkpoint
restart this process is inverted, and the root processor reads the
maps (and scalars) from disk and scatters them to the other
processors. In this way the checkpoint data volume and distribution is
independent of the number of processors used, and a restarted job is
not required to use the same number of processors as the original.  Figure
\ref{mapflow} is an overview of the map making process.

\subsection{Data Distribution}
 The memory for both the time domain data and the pixel domain data are
distributed over processors.  The user has options regarding how the time
domain data will be distributed in ways that allow for CPU and memory
resource optimization depending on the details of the problem being
solved and the architecture of the available hardware.  MADmap has several options that allow the user to tune the
total memory consumption in compromises between memory and CPU
resources.

MADmap uses the MPI library to enable the use of distributed parallel
computing resources.  MPI uses a single program multiple data parallel
model.  The computational load scales with the number of time samples
assigned to each processor.  In order to load balance computation we
attempt to load balance total sample count.  
There are three memory modes in which MADmap can be run.  In the high
memory mode, the full detector pointing is held in memory
concurrently.  The other data objects which are required in high
memory mode are the noise filters.  In low memory mode, the compressed
telescope pointing and the noise filters are the only persistent time domain data
objects.  In very low memory mode, only the compressed telescope
pointing remains in memory and the memory requirement is independent of the detector count.

MADmap distributes its data by dividing the time stream data over the
processors.  Then the pixels are distributed so that each processor
stores the pixels observed in the sections of the detector time
streams that are assigned to it.  MADmap is enabled to use three
different data distributions: concatenated mode, stacked mode, and 
multi-stacked mode.  In
the concatenated distribution each detector time stream is
concatenated with all the others and the resulting ``super'' time stream
is evenly divided among the processors.  In the stacked distribution
each processor analyzes data from all detectors, and the experiment's
run time is divided evenly over the processors.  The multi-stacked 
distribution is a compromise between the other two, and 
provides the option to stride the detectors over the
processors.  That is to say that if the stride was $s$, then each
processor would analyze the data from the fraction $1/s$ of
the detectors, and the experiment's run time is evenly divided among
processors who have a common remainder when divided by $s$.

\begin{figure*}[htbp]
\begin{center}
\includegraphics[scale=1.0]{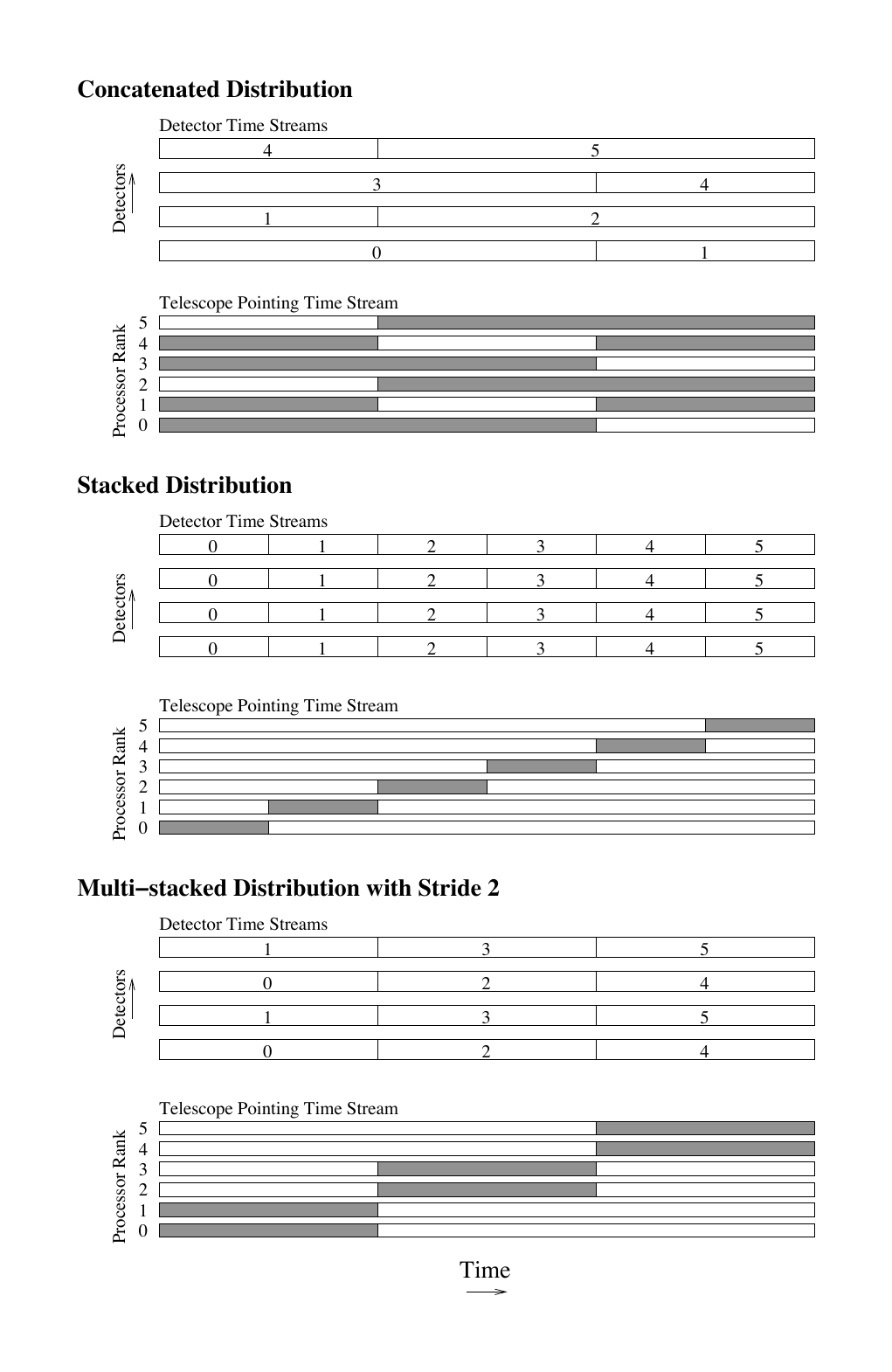}
\caption{This is a graphical representation of the data distribution
  for the detector time streams and the telescope pointing time
  stream.  The cartoon represents the distribution of four detector
  time streams over six processors in the three different modes of
  operation.  The numbers labeling the figure all represent processor
  ranks.  Note that the detector time streams are distributed over the
  processors, where as the telescope pointing is only fully
  distributed in the stacked mode.  For this reason we can show the
  distribution of the detector time streams over all of the processors
  in one image but the telescope pointing data distribution must be
  shown for each processor individually where gray indicates telescope
  pointing that is stored on a given processor.
  \label{labelDataDistribution}
}
\end{center}
\end{figure*}

\begin{figure*}[htbp]
\begin{center}
\includegraphics[scale=0.6]{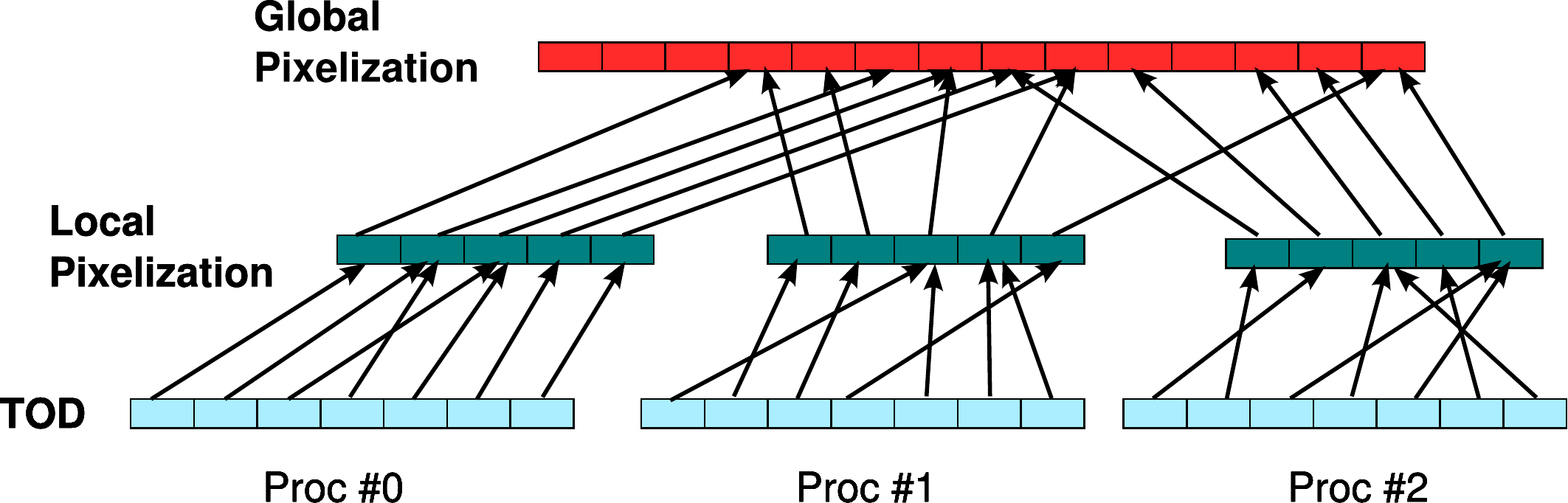}
\caption{This is a graphical representation of the distribution of pixel
domain data between processors.  Each processor has a set of time samples
and the minimal set of ``local pixels'' needed for the pointing in those
samples.  These local pixels have a further mapping to global pixels on
the sphere.
\label{pixeldist}
}
\end{center}
\end{figure*}

Figure \ref{labelDataDistribution} is a graphical representation of
the memory distribution for the detector time streams and telescope
pointing.  Figure \ref{pixeldist} shows a toy mapping of each processors'
locally observed pixels to the global pixel space.  The pixel data that
is stored in the memory of each processor corresponds to those pixels 
which were observed by the detector samples within the time streams 
analyzed by the processor.  This means that the distribution of the pixel 
data is strongly dependent on the scanning strategy of the experiment 
and the relative locations of the detectors on the focal plane in 
addition to the effects of the different modes of time stream distribution 
and the number of processors used to analyze the data.

The concatenated distribution has several advantages over the
unstrided stacked mode, but there are some very critical advantages to
the stacked mode that make it quite useful.  The multi-stacked
distribution is often a good compromise that will allow the problem to
be tackled with the computational tools available given a good choice
of stride.  In the unstrided stacked mode all of the the
detector-specific noise filters are required to be stored on each
processor, compared with very few in the concatenated mode.  The
concatenated distribution can be used to simultaneously analyze data
from different experiments, or data sets that have large gaps in time
where no data are taken.  In the case of run time simulation of
correlated noise, the concatenated mode allows for load balanced
generation of correlated noise for longer intervals of time on a each
processor than in the unstrided stacked mode.  In this case the use of
the concatenated mode provides the ability to scale this calculation
to larger numbers of processors.  The unstrided stacked distribution
is advantageous because this is optimal for the storage and
computational requirements of expanding the compressed pointing data
that is shared by all the detectors for each time sample.

\subsection{Weighting} \label{sec:weighting}

If the instrumental noise is a least piece-wise stationary, $N$
will be a block-diagonal Toeplitz matrix, with each block
corresponding to one of the stationary pieces. A Toeplitz matrix is defined so that each row is shifted by one entry from the one above. For an $n\times n$ Toeplitz matrix, 
\begin{equation}
T_{i,j} = T_{i-1,j-1} \ \forall\  i\ \&\ j \in \{1,\ldots,n-1\}\;.
\end{equation}
Moreover, for a typical
experiment each of the diagonal blocks will be banded. 
Generally, the inverse of a Toeplitz matrix is not Toeplitz, however
for banded matrices this is nearly true with potential departures seen
only at the first and last rows and columns of the diagonal block \citep{bib:stompor02}. Moreover, the inverse of a banded-matrix
is also, to a good approximation, banded with a band-width which needs to
be appropriately tuned. The MADmap algorithm assumes
therefore that $N^{-1}$ is a block-diagonal, banded Toeplitz
matrix.
\begin{equation}
N^{-1}_{t,t'} = \left\{ \begin{array}{ll}
	        f(|t-t'|) & {\rm if} \;\;\; |t-t'| < {\rm n_c} \\
                0 & {\rm otherwise}
	                 \end{array}
                \right.
\end{equation}
We note that though it is a very good approximation, it is still an
approximation; however, at worst it can only compromise the optimality
of the MADmap map without biasing it.

The method used by MADmap for convolution with a filter is similar the
the ``overlap and add method'' outlined in Numerical Recipes  \citep{bib:press92}.
 The method given here is slightly
different and requires fewer computations by avoiding two correlation
length additions for each FFT computed.  The method presented also
gives an algorithm for choosing the FFT length which is optimal 
in terms of number of operations required.  

The discrete Fourier transform operator diagonalizes circulant
matrices, and it is well known that the fast Fourier transform
algorithm of Cooley and Tukey 
\citep{bib:cooley65} can be used to quickly compute the action of a
circulant matrix on a vector.  This algorithm has a computational
complexity of $O({\rm n_f} \log({\rm n_f}))$, where ${\rm n_f}$ is the length of the vector
and size of the matrix.  A circulant matrix is a special type of
Toeplitz matrix with the additional constraint of periodic 
boundary conditions \citep[e.g.,][]{bib:golubLoan96}:
\begin{eqnarray}
C_{i,j} = C_{i-1,j-1}\ \forall\ i\ \&\ j \in \{1,\ldots,{\rm n}-1\}\\
C_{0,j} = C_{{\rm n}-1,j-1}\ \forall\ j \in \{1,\ldots,{\rm n}-1\}\\
C_{i,0} = C_{i-1,{\rm n}-1}\ \forall\ i \in \{1,\ldots,{\rm n}-1\}
\end{eqnarray}
The inverse time-time noise correlation matrix, $N^{-1}$ is block
diagonal and these blocks are themselves band diagonal and Toeplitz.
Our objective is to use circulant matrix multiplies as a building
block from which we can compute the action of $N^{-1}$.  We can
treat each diagonal block of $N^{-1}$ as an independent problem, so
let us simply consider the problem of multiplying a band diagonal
Toeplitz matrix with a bandwidth of ${\rm n_c}$ where 
\begin{eqnarray}
B_{i,j} = B_{i-1, j-1}\ \forall\ i\ \&\ j \in \{1,\ldots,{\rm n_c}\}\\
B_{i,j} = 0\ \forall\ i\ \&\ j\ \mathrm{s.t.}\ |i-j| > {\rm n_c}
\end{eqnarray}
We can approximate $B$ with a circulant matrix $C$ which differs from $B$ 
only in the upper right and lower left corners of the matrix 
\begin{equation}
B_{i,j} = C_{i+b,j+b}\ \forall\ i\ \&\ j\ \mathrm{s.t.}\ |i-j| < {\rm n}-{\rm n_c}
\end{equation}
In order to compute the action of $B$ on a vector by using a
circulant matrix $C$, we construct $C$ to be two bandwidths larger than $B$
and pad the vector to be multiplied with a bandwidth of zeros at the
beginning and end.  The vector that results from the multiplication of
$C$ by the zero-padded vector will be the correct solution to the
original problem once a bandwidth has been trimmed from the 
beginning and end of the resulting vector.  

The technique above scales as $O({\rm n} \log({\rm n}))$ where here ${\rm n}$ is the
length of a stationary period.  In the case where ${\rm n} \gg {\rm n_c}$ we can do
even better than this.  As shown above, each time we do a circulant
matrix multiply in place of a banded Toeplitz matrix multiply there is
corrupted data within one bandwidth of either end of the resulting
vector.  We can use this fact to break the problem up into two
sub-problems where two circulant matrix multiplies are performed, each
of ${\rm n}/2 + 2 {\rm n_c}$ length, and each producing correct results for ${\rm n}/2$
elements.  The vectors multiplied would be padded with zeros on the
two ends of the original vector, as before, but the end of the first
vector would be extended with the beginning of the second half of the
original, and the beginning of the second vector would be padded with
the end of the first half of the original vector.  This idea can be
extended further and the problem can be broken up into as many
sub-problems as we would like, to the limit where each circulant
matrix multiply produces a single uncorrupted element for the output
vector.  The complexity of the suggested algorithm can be determined
as follows: the cost of each FFT is ${\rm n_f}\log({\rm n_f})$, and the number of
uncorrupted elements determined by an FFT is ${\rm n_f} - 2 {\rm n_c}$.  In order to
calculate ${\rm n_t}$ uncorrupted elements there must be $\lceil
{\rm n_t}/({\rm n_f}-2 {\rm n_c})\rceil$ FFT's performed, so the computational
complexity is
\begin{equation}\label{eq:FFTcomplexity}
O\left(\left\lceil \frac{{\rm n_t}}{{\rm n_f}-2 {\rm n_c}}\right\rceil {\rm n_f} \log_2({\rm n_f})\right).
\end{equation}
To a certain extent ${\rm n_f}$ is a free parameter and we can choose it to
be optimal.  This is done by minimizing the complexity.  For
simplicity of analysis we consider the function $h({\rm n_f})$:
\begin{eqnarray}
h({\rm n_f}) =\frac{{\rm n_f}  \ln({\rm n_f})}{{\rm n_f}-2 {\rm n_c}} \label{eq:FFTcomplexityFunction}\\
h'({\rm n_f}) = \frac{1+\ln({\rm n_f})}{{\rm n_f}-2{\rm n_c}} - \frac{{\rm n_f} \ln({\rm n_f})}{({\rm n_f}-2{\rm n_c})^2}\\
h''({\rm n_f}) = \frac{2 {\rm n_f} \ln({\rm n_f})}{({\rm n_f}-2{\rm n_c})^3}-\frac{2(1+\ln({\rm n_f}))}{({\rm n_f}-2{\rm n_c})^2}+\frac{1}{{\rm n_f}({\rm n_f}-2{\rm n_c})}
\end{eqnarray}
Equation~(\ref{eq:FFTcomplexity}) and Equation~(\ref{eq:FFTcomplexityFunction})
differ in that Equation~(\ref{eq:FFTcomplexity}) includes a
discretization and a linear factor of ${\rm n_t}/\ln(2)$, which determines
the discretization.  We will revisit the discrete nature of the
problem later, as there is another discretization to consider which
impacts performance: the radix of the FFT.  Note as well,
that we only consider the cases where ${\rm n_f} > 2 {\rm n_c}$, as all values
determined by the FFT would be corrupted by the boundary conditions in
cases where ${\rm n_f} \le 2 {\rm n_c}$.  Setting $h'({\rm n_f}) = 0$ gives the following 
relationship between ${\rm n_f}$ and ${\rm n_c}$:
\begin{equation}
{\rm n_f} - 2{\rm n_c}(1+\ln({\rm n_f})) = 0
\end{equation}
which, unfortunately, does not have an analytical solution for ${\rm n_f}$ in
terms of ${\rm n_c}$.  We can solve for ${\rm n_c}$, however, and check that
$h''({\rm n_f}) > 0$ at this critical point, and this is shown in Appendix~\ref{Appendix:optimalFFT}.  

In practice, radix two FFT's are significantly faster than FFT's of
vectors of a length with a more elaborate factorization than simply a
power of two.  There is also no sense in choosing an FFT length that is
more than twice the length of the longest stationary period to which
it will be applied.  As a result, we would like to chose the
smallest power of two larger than the longest stationary period.  If
this value of ${\rm n_f}$ gives an $h'({\rm n_f})$ which is positive, then decrease
${\rm n_f}$ by a factor of two until $h'({\rm n_f}/2) < 0$ or ${\rm n_f}/2 < 2 {\rm n_c}$. This value
of ${\rm n_f}$ is the length of the FFT used by MADmap in the case where the user 
has not specified a length by way of an environment variable\footnote{The FFT length environment variable: {\tt MM\_FFT\_LENGTH}}.  

MADmap uses the FFTW library by default \citep{bib:frigo05}, 
however at compile time it is also possible to
specify the ACML library to do FFT's.  The ACML library is
significantly faster than FFTW on AMD's Opteron platform.  Future
extensions to MADmap will allow the use the Intel Math Kernel library
and IBM's libmass which are other common high performance math
libraries which include FFT functionality.

\subsection{Compressing the pointing matrix}

The pointing matrix is at the core of our time stream model
as outlined in Equation~(\ref{eDSN}).  This matrix determines
how a time stream of data observed by a detector relates to the
discretized signal $z$.  In its simplest form the pointing
matrix can have one non-zero per row with value unity.  The column
assigned this unity value is determined by the pixel index of the element 
in the map that is being observed by a detector at the time sample 
associated with the row.  More sophisticated pointing 
matrices result when there is polarization sensitivity, or when 
the pointing matrix includes information about the 
detector beam sensitivity pattern.  Each row of the pointing 
matrix determines the linear combination of the discrete signals being 
measured that were observed in a particular time sample.  

As discussed in the introduction, the compression of the pointing 
matrix is critical to reducing the memory requirement,
especially for analysis of data collected by contemporary CMB
experiments with large numbers of detectors on a single focal plane.
The calculation of the ``right hand side'' (RHS): $ A^T N^{-1} \gamma$, uses
$\gamma$, which is a length ${\rm n_t}$ vector, but this vector is used only once
with a single pass through the time stream, so there is no need to
store all of $\gamma$ in memory simultaneously.  The data can be read into
small buffers, reduced to pixel domain data, and then overwritten
with the next buffer of data.  

The pointing matrix $A$, however, appears in the right hand
side, and twice in the linear system inverted, which means that $A$ is
used on each iteration of our PCG solver.  The simplest solution to
this problem is the ``high memory mode'' of
MADmap, where the sparse representation of $A$ is stored concurrently
in memory (in a distributed fashion) for the entire execution of the
application.  This method is fast, but can be overwhelmingly memory
intensive, as the memory requirement scales as $O({\rm n_t})$.

One method of averting this problem is to attempt to read the pointing
matrix into memory from disk on each iteration of the PCG, but for
almost every system configuration this will result in an I/O bound
problem, and leave the processors idle while they wait on the disk
subsystem.  We have not yet attempted to use the asynchronous I/O
options afforded by the MPI-2 standard, but a better solution exists 
which avoids repeated reading altogether while requiring a more modest
memory footprint than is required for holding all of the detector pointing 
in memory simultaneously.

To overcome this problem, we make use of the fact that our pointing 
matrix is derived from data which is smaller than data
required for the sparse representation.  The compression of the pointing 
matrix is experiment and signal specific, but in very broad
terms, MADmap can work with any representation of compressed pointing
that is sampled at a constant rate with a fixed number of double
precision floating point numbers per sample.  These data can be
ingested by the GCP library which
will produce detector-sampled pointing in row compressed sparse form
(column index and weight pairs).  

The pointing matrix for CMB experiments is often derived from
telemetry data which measures the direction in which some fiducial
center of the focal plane is pointed at a rate that is significantly
lower than the detector sample rate.  This is true for both real data
sets and often for simulated data as well.  The sparse representation
of $A$ is derived by an interpolation of the bore-sight pointing to a
detector sample rate, and a rotation of the bore-sight pointing to
each detector's relative offset.  This location is then discretized
into a pixel index and assigned a weight, which usually differs from
unity if the detector is polarization sensitive and the calculation
of this weight is experiment specific.

In the past, when the number of detectors being analyzed
simultaneously was more modest, the pointing solution for each
detector would be stored on disk at the detector sample rate.  This
would sometimes be stored as ordinates (e.g., Euler angles) or as a
list of pixel indices (with or without weights depending on the
experiment's capacity to measure a polarized signal).  Reading these
detector specific pointing sampled at the detector sample rate can be
quite costly, but is still a possible option in MADmap.

\subsection{Pixel indexing and distributed signal vectors}

Regardless of how the sparse pointing matrix is obtained,
once it is constructed in terms of the global pixel indexing scheme it
must be re-indexed to match the distributed local pixel indexing
scheme.  Every pixelation of the sky has some intrinsic mapping
between an index value and a position on the sky, and likewise, any
discretized signal will have a mapping between index and the element
of the discrete signal.  There are actually three different indexing
schemes used by MADmap: the global indexing scheme which is intrinsic
to the discretization of the signal, the observed indexing scheme
which includes only those index elements which are observed in any
processor's data, and the local indexing scheme which includes only
those index elements which were observed within the local processor's
data.

In order to distribute the signal vectors over the processors, each
processor stores only those signal elements that are observed during
the samples that are assigned to it, and all local signal vectors are
indexed with the local scheme.  MADmap stores two vectors that are
used for mapping between indexing schemes.  These are both the length
of the number of locally observed pixels, and map from local index to
observed index and from local index to global index.  In order to use
these vectors to map from global or observed index to local index, a
binary search of the appropriate vector is done.  This re-indexing is
done every time the sparse pointing matrix is constructed
from compressed pointing data.  This requires a binary search of the
locally observed pixel index for each time sample which has a
computational complexity of $O({\rm n_i n_t} \log({\rm n_z}))$ operations.  
These are integer operations and do not contribute to the total flop count \emph{per se},
but do consume clock cycles none the less.  

In order to construct the index remapping vectors without ever
allocating a full ${\rm n_z}$ length indexing vector, some bit operations are
used.  A bit array is allocated that has one bit for every global
pixel index (${\rm n_z}$ bits).  The elements of this array are set to one
or zero depending on if the index is observed within the local pointing 
matrix.  This array can be used to construct the mapping
from local index to global index.  This array is then collectively
reduced using \verb!MPI_AllReduce()! with the \verb!MPI_BOR! operator
after which each processor has a bit vector that tells which pixels
were observed within the data on all processors, and this bit array
can be used to construct the mapping vector from local to observed 
pixels.  

\subsection{Communication patterns}
Given the size of CMB data sets, MADmap is intended for parallel
machines, from small clusters to the most massively parallel systems
extant. Both time-ordered and pixel-domain data are distributed over
the processors, and MADmap uses the standard Message Passing Interface
(MPI) for inter-processor communication. Since the available memory
will vary hugely depending on the number of processors being used to
solve a particular problem (from tens to tens of thousands at the time
of writing, with hundreds of thousands on the immediate horizon) and
the available memory per processor (typically from 256MB to 4GB),
MADmap provides a number of options to trade additional calculation
for reduced memory.

Fundamentally, there are three communication operations in MADmap.  
The first comes up in the case where each processor has
computed the contribution to a map vector from the time samples
assigned to it in the computation of the operation of the pointing 
matrix.  The sub-maps on each processor must be summed with
the sub-maps on all other processors.  Note that each processor
contains a subset of the observed pixels that is overlapping with the
subset of observed pixels on other processors.  This computation, in
the end, is computed with a series of \verb!MPI_AllReduce()!  calls.
Each call is done on a fraction of the entire observed map, and each
processor fills the reduction vector with the value it computed, or
zero if the pixel was not observed in the processor's time stream.
After the reduction is completed, the results that are pertinent to
the processor are stored, and the vector is overwritten with the next
buffer to be reduced.  This operation is called
\verb!MM_SomeReduce()!.  This is, by far, the most costly communication
operation.  It scales like $O({\rm n_z} \log({\rm n_{proc}}))$ where ${\rm n_z}$ is the
number of observed pixels, and ${\rm n_{proc}}$ is the number of processors.
This must be done once on each iteration of the PCG algorithm.

The other two communication operations depend on the concept of
``primary pixels.''  In MPI, each processor available to an application
is assigned a rank which is an integer in the set $\{0,1,\ldots,{\rm n_{proc}}-1\}$ 
so that every processor has a unique rank and there is a sequential ordering of the 
processors.  A processor's primary pixels are defined to be
those pixels which are observed within the time stream distributed to
the processor, and are not observed within the time stream distributed
to any lower rank processor.  
This determination defines a disjoint
division of the pixels over the processors that is comprised of a
subset of the fundamental distribution of the pixels over the
processors.  To determine the primary pixels, an \verb!MM_SomeReduce()!
is called on a vector that is set to be the processor's rank if the
pixel is observed by the processor or the total number of processors
in the case where the pixel is not observed within the processor's
time stream, and the reduction operator is \verb!MPI_MIN!.  After the
call to \verb!MM_SomeReduce()!, the reduced vector can be used to
determine as primary pixels all of the pixels that are set to the value of
the processor's rank.

The second operation is very similar to the first, except that it is
used only in the writing of maps to disk.  When writing a distributed
map vector to disk there are a series of calls to \verb!MPI_Reduce()!.
This sequence of reductions effectively constitute single reduction operation on a map of all observed pixels while limiting 
the size of the vector used in the \verb!MPI_Reduce()! call and thereby mitigating the memory requirement.  
In this operation each processor sets the map value to its stored value if the pixel is a primary pixel,
and sets it to zero if the pixel is not a primary pixel of the processor. 
In this way the root processor collects the map
and writes the buffer to disk between each call to
\verb!MPI_Reduce()!.

The final operation is the computation of the dot product of two
distributed map vectors.  This is done by computing the local
contribution to the dot product by primary pixels on the processor,
and then an \verb!MPI_AllReduce()! is called on the one double
precision floating point element to evaluate the complete dot product
combining calculations from each processor.  Because this requires a
single call to \verb!MPI_AllReduce()!  on a single element for each
dot product, this communication cost does not account for a
significant amount of the run time.

\subsection{Memory Use and Computational Complexity}
MADmap's computational work can be segmented into several distinct
operations.  The dimensions of the problem to be solved determine the
relative cost of the operations, but we will discuss the operations
that typically are the most time consuming.  Those operations which
are done on each iteration of the PCG tend to dominate the run time
because the factor of the number of iterations is usually between one
and two orders of magnitude.  The dominant term in the overall
computational complexity of MADmap in all modes of operation is $O({\rm n_i  n_t} \log( {\rm n_c} ))$, which is determined by the computation of the
FFT's, and this computation constitutes a significant portion of the
run time.  A portion of the run time is spent constructing and acting
with the pointing matrix and, in low memory mode, these
operations have an operational complexity of $O({\rm n_i  n_t n_{nz}}( 1 + \log( {\rm n_z}))$ with a significant pre-factor.  This computation includes
the expansion of the compressed pointing into the sparse pointing 
matrix, the re-indexing of the pixels to distributed
indexing, and multiplying by the sparse re-indexed pointing matrix.  

The complexity measures given below are distributed over the
processors and MADmap has been scaled to run on tens of thousands of
cores.  MADmap is load balanced computationally and the memory
distribution is balanced in all dimensions except ${\rm n_z}$ which is,
nonetheless, distributed.  MADmap has three different modes of
operation which reduce the memory requirement at the expense of more
computations.  The operational complexity of the high memory mode is
\begin{equation}
{\rm c_h} = O({\rm n_i} {\rm n_t} \log({\rm n_c})),
\end{equation}
and the memory requirement is 
\begin{equation}
{\rm m_h} = O({\rm n_z} + {\rm n_t} + {\rm n_{proc}} {\rm m_n}).
\end{equation}
The operational complexity of the low memory mode is 
\begin{equation}
{\rm c_l} = O({\rm n_i} {\rm n_t} ( \log({\rm n_c}) + n_{nz} \log({\rm n_z}) )
\end{equation}
and the memory requirement is 
\begin{equation}
{\rm m_l} = O({\rm n_z} + {\rm r_p} \Delta t + {\rm n_{proc}} {\rm m_n}).
\end{equation}
The operational complexity of the extremely low memory mode is 
\begin{equation}
{\rm c_e} = O({\rm n_i} ({\rm n_t} ( \log({\rm n_c}) + n_{nz} \log({\rm n_z}) ) + {\rm n_b} {\rm n_c} \log({\rm n_c})))
\end{equation}
and a memory requirement of 
\begin{equation}
{\rm m_e} = O({\rm n_z} + {\rm r_p} \Delta t).
\end{equation}
The variable ${\rm m_n}$ is the per processor memory requirement for the noise
filters and depends on the data distribution as given by the 
variables ${\rm m_{ns}}$ and ${\rm m_{nc}}$ defined in Section \ref{sec:diskUsage}.  
Note that low memory mode footprint is nearly invariant with the
number of detectors and the footprint of the extremely low memory mode
is completely invariant with the number of detectors.  Next generation
CMB experiments will operate with large numbers of detectors sampled
at a high rate so the memory savings afforded by low memory mode will
be critical to the tractability of the maximum-likelihood data
analysis of these next generation experiments.

\subsection{Disk Usage} \label{sec:diskUsage}
The disk is accessed for five purposes, three of them reading:
detector time streams, pointing information, and noise filters; and
two of them writing: checkpointing, and the final maps.  There are
many options which allow most of this disk access to be minimized.  In
some cases the time stream data are simulated at run time using the M3 on-the-fly simulation capabilities,
and require only a small amount of input data from disk (usually in the
form of maps).  The pointing information can take on a wide variety of
forms.  Noise filters can be input parametrically, and in this case
the filters are calculated at run time without reading from disk.
Checkpointing is optional, and the frequency can be chosen, but by
default occurs after every 20 iterations.  The output of the final
maps is a relatively small amount of data, and only one processor
writes the final maps to disk.  

The detector time stream samples are distributed in a balanced way
over the processors independent of the data distribution mode chosen.
Each processor reads or simulates a fraction of the total volume of
the detector time samples divided by the number of processors.  Recall
that there are three data distribution modes in MADmap: concatenated,
stacked and multi-stacked.  The choice of distribution has an impact
on the amount of telescope pointing data read, and the number of noise
filters read and stored, but does not impact the volume of detector
data required by each processor.  If the time stream
data are read from disk, then every processor reads 
\begin{equation}
{\rm d_t} = O\left({\rm \frac{n_t}{n_{proc}}}\right)
\end{equation} 
samples which are typically stored in eight byte precision.

The amount of pointing data to be read from disk depends on how the
MADmap job was configured.  In the minimal case, MADmap is run in the
stacked data time stream distribution, and telescope centroid
pointing sampled at a slow rate is used and interpolated to detector
pointing at run time.  The amount of reading required in this case is
the number of telescope centroid pointing samples divided by the
number of processors.  
The use of the concatenated data distribution
requires an additional factor of the number of focal plane detectors of
extra reading for centroid pointing since different processors are analyzing data from the same time period.  The telescope centroid
pointing data in the concatenated data distribution are not balanced and different processors will require the same pointing data.
When using telescope centroid pointing it is usually optimal to
use the stacked data distribution. In the typical case of reading
telescope centroid pointing and using the stacked distribution MADmap
reads 
\begin{equation}
{\rm d_{ps}} = O\left({\rm \frac{ n_t r_p}{n_d r_d n_{proc}}}\right)
\end{equation}
elements of the telescope pointing.  Each
element of telescope pointing is typically between three and six
numbers stored in eight byte precision.  Using the concatenated
distribution requires another factor of ${\rm n_d}$ on each processor 
bringing the number of elements to 
\begin{equation}
{\rm d_{pc}} = O\left({\rm \frac{ n_t r_p}{r_d n_{proc}}}\right).
\end{equation}
If individual detector pointing is read the number of elements
required is 
\begin{equation}
{\rm d_{pd}} = O\left({\rm \frac{ n_t}{n_{proc}}}\right).
\end{equation}

The potential hazard of the stacked data distribution lies in the
reading and memory requirement associated with the noise filters.  In
the concatenated distribution each processor has to load and store
only order one noise filter requiring 
\begin{equation}
{\rm d_{nc}} = {\rm m_{nc}} = O({\rm n_c})
\end{equation}
elements read from disk stored in eight byte precision by each
processor.  In the stacked distribution each processor loads and
stores a noise filter for every detector requiring
\begin{equation}
{\rm d_{ns}} = {\rm m_{nc}} = O({\rm n_d n_c})
\end{equation}
elements stored and read from disk.  To compromise between these noise
filter requirements and the telescope centroid pointing requirement the
multi-stacked distribution is advised.  An exceptional case for the
filter memory requirement occurs if the same noise filter is used for
every stationary interval analyzed, and in this case MADmap stores
just one filter, regardless of the data distribution.  This exception
is most useful for simulated data.

\section{Using MADmap} \label{sec:using}
The MADmap algorithm, its implementation, as well as the structure of the associated
data abstraction (M3) and compressed pointing (GCP) libraries have all been designed to
ensure maximal flexibility and broad applicability of the software to CMB data analysis
and more generally to estimation problems.  The range of problems that the MADmap
software can be directly applied to is indeed very wide, and
significant effort has been undertaken to ensure that the generality
in the problem formulation and algorithm choices do not negatively impact the
computational efficiency of the software. We will illustrate
the efficiency issue in the next section studying two specific
applications and compare it with some other similar codes in
Section~\ref{sect:comparison}.  In this section we describe a set of
example problems to which MADmap can be directly applied. We include
here only the kinds of runs which have actually been performed and
their results validated on either simulated or real data.  

MADmap's most notable flexibility is its ability to ingest 
and process an arbitrary pointing matrix. This is true at least on the 
algorithmic level, however the run-time and memory requirements will 
depend on the sparsity and structure of the pointing matrix, and these requirements will clearly set some 
constraints on the practical level. Nonetheless, MADmap permits performing 
runs which 
not only produce the sky signal maps, but also extract the unwanted
parasitic contributions typical of many experimental data, as long as they 
can be expressed within the linear response model of Equation~\ref{eDSN}.
The examples given previously demonstrate that those include cases of real 
practical interest.

\noindent(1) {\it Non-parametric synchronous signals.}
In many applications there exists a parasitic signal of an instrumental or environmental
origin which can be thought of as a unique function of some parameter.
We will refer to those as signals synchronous with the respective parameter.
Examples of such signals include the previously mentioned primary-mirror-chop-synchronous parasitic signal observed in the MAXIMA data, as well as ground or atmospheric 
pick up usually seen in ground-based experiments,~\citep[e.g.,][]{bib:kuo04}. 
If we denote such a parameter or set of those by 
$\psi$ we can update Equation~\ref{eDSN} to read,
\begin{equation}
\gamma = \nu + \zeta + \psi = \nu + A z + B y
\label{eDSN1}
\end{equation}
where $B$ is a `pointing' matrix of the parasitic signal, and
$y$ is a discretization of its amplitude. In the specific example of
the primary-chop-synchronous parasitic signal, $y$ may correspond to an orientation
angle of the primary with respect to the gondola frame and $B_{t,p}$ is non-zero and
equal to unity for all the time samples, $t$, for which the primary orientation fell in between
an interval, $\left[y_p, y_p+\Delta\,y_p\right]$, defining the non-overlapping
discretization (one dimensional 'pixelation') of the possible angles, $y$.

On replacing,
\begin{equation}
z \rightarrow 
\left[
\begin{array}{c}
z\\
y
\end{array}
\right],
\ \ \ \ \ \ 
A \rightarrow 
\left[
\begin{array}{c c}
A & B
\end{array}
\right],
\label{eqn:PmatExt}
\end{equation}
in Equation~\ref{eDSN} we see that both Equation~\ref{eDSN} and Equation~\ref{eDSN1} are 
clearly of the same form 
and therefore can be solved using the same algorithm as implemented in
MADmap, provided that suitable input data properly describing the
full pointing matrix of the problem are available.

\noindent(2) {\it Parametric synchronous signals.}

In some applications parasitic signal may not be synchronous with any easily
identifiable external parameter but there may exist a linear parametric model
which can describe the time-dependence of the signal, i.e., $\psi_t \equiv
\psi\left( t; \left\{b_i\right\}\right)$ and $\psi = C b$. We note that the last
equality may be in fact only an approximation, e.g., a result of linearization 
procedure around the anticipated values of the parasitic signal parameters, even 
if the latter are inherently non-linear. In those cases, the validity of the approach 
may need to be evaluated \textit{a posteriori}.

On adding such a term to Equation~\ref{eDSN}, we again arrive at an expression 
similar to Equation~\ref{eDSN1}, which in turn is equivalent to Equation~\ref{eDSN} with
appropriately redefined pointing matrix and estimate vector, Equation~\ref{eqn:PmatExt}.

We note that such an option is of great current interest.
This is because the parasitic signal due to the rotating polarizers, such
as e.g., half-wave plate. The signal of this sort is expected to be one of
the major systematics the polarized experiments using this kind of technology, and 
accurately removing this systematic is crucial to achieving sensitivity levels required for
a successful B-mode polarization detection. It has been shown~\citep{bib:johnson07} 
that in cases of practical interest such a signal can be modeled with a 
couple of dozens of parameters and that a linear model can be indeed sufficient for 
such a purpose. Specifically, the particular model can be written down as,
\begin{eqnarray}
\psi_{\rm HWP}\left(t\right) & = & \sum_{k=1}^8\,\left(b_k+t\,b_{k+8}\right)\,\cos k\omega_t
\ \ \ \ \ \ \nonumber \\
& + & \left(b_{k+16}+t\,b_{k+24}\right)\,\sin k\omega_t,
\end{eqnarray}
where $\omega_t$ denotes a known position angle of the polarizer. 
Thus for any time $t$, we recover,
\begin{equation}
C_{t,i} = \left\{
\begin{array}{l l}
{\displaystyle \cos i\omega_t} & {\displaystyle i = 1,...,8;}\\
{\displaystyle t\cos (i - 8)\omega_t} & {\displaystyle i = 9,...,16;}\\
{\displaystyle \sin (i - 16)\omega_t} & {\displaystyle i = 17,...,24;}\\
{\displaystyle t\sin (i - 24)\omega_t} & {\displaystyle i = 25,...,32;}
\end{array}
\right.
\label{eqn:HWPmat}
\end{equation}
where $C$ complements the sky signal pointing matrix as in Equation~\ref{eqn:PmatExt}. The map-making
solver will then estimate simultaneously both the sky signal ($z$) and the HWP parasitic
parameters ($b$).

We note that, in the case of the rotating polarimeter, the parasitic signal
is indeed synchronous with the polarizer orientation and could also be subtracted 
using the approach described in case (1) above. Applying the parametric model 
as described here, however, 
allows us to reduce the number of extra degrees of freedom introduced
to the problem and therefore to achieve higher precision of the recovered
final maps of the sky. It also permits for the avoidance of discretization effects, and these are particularly
important for cases involving rapidly varying systematics.
We also point out that the block of the pointing matrix, as defined in Equation~\ref{eqn:HWPmat},
is dense, and therefore poses a particular challenge to the standard way of implementing
the pointing and unpointing operators discussed earlier.  

\noindent(3) {\it 'Destriping' runs.}
In some particular applications the time-domain noise can be 
effectively modeled as white noise plus a series of random offsets assigned
to some predefined time intervals. This was first discussed in the context
of Planck-like experiments~\citep{bib:janssen96} and gave rise to so-called 
destriping algorithms~\citep{bib:delabrouille98,bib:maino99,bib:keihanen04}. It has 
been shown that in the Planck case the destriper achieves a competitive 
precision at a fraction of the numerical load~\citep{bib:poutanen06,bib:ashdown07a,bib:stompor04}. 
The approximate time-order data model in this case reads
\begin{equation}
\gamma_t = \nu + \zeta + \omega = \nu + A z + B o
\end{equation}
where $\nu$ is uncorrelated, piecewise-stationary noise, $\omega_t$
is an offset added to a sample $t$, and $o_p$ is an offset common
to all samples in a time interval $p$. $B$ is an offset
pointing matrix, essentially defining the time intervals associated
with each of the offsets denoted as $o_p$. Using both the pointing 
matrix and the solution (map) vector extended as in Equation~\ref{eqn:PmatExt}
MADmap can therefore simultaneously estimate the sky signal and the offset
amplitudes,
assuming that the used time intervals conditions do not lead to
a singular system of equations.
 We can further assume that the noise is not merely 
stationary, but \emph{white}, and thus that $N$ is diagonal,
which is a basic assumption in the standard 
destripers~\citep{bib:keihanen04}.
In spite of its more complex pointing matrix, 
this kind of run will benefit considerably from the
fact that no FFT is needed for performing any of the noise kernel
convolutions, and this results in a significantly shorter run-time
when compared to the run incorporating the more complex time-stream model.
As mentioned before, at least in some specific cases, the speed-up may
indeed be achieved at no substantial loss of precision in the final
sky maps. We note however that though such MADmap run would indeed be
significantly faster, than a full MADmap run with a complete noise model,
it would not reach the speed typical of custom-made
destripers. Nevertheless, the attractive feature is that MADmap provides 
in a single package both functionalities as well as entire spectrum of
intermediate options.

\noindent(4) {\it Multi-resolution map-making}
As MADmap does not interpret or make any assumptions about the pixel signal
and/or pixelation schemes, it straightforwardly permits for mixing sky
pixels of different resolutions, for instance, enabling a higher resolution for sky
areas with particularly high signal-to-noise ratio. It also accepts
different resolutions for different Stokes parameters. The latter fact is
again particularly interesting in the light of forthcoming polarization
experiments based on total power detectors. In such cases the measured 
signal is a linear combination of all three Stokes parameters. Map-making
therefore attempts to unscramble it to estimate all three of them 
simultaneously. For the CMB, the power contained in the total intensity mode, 
$I$, is significantly higher than that in the polarization. That
has two consequences. First, the total intensity map could be pixelized
with high resolution pixels retaining the same signal-to-noise on the pixel
scale as for the polarization. Moreover, given that sky signal is assumed
to be constant on the pixel scale, any departure from such assumptions
(e.g., due to residual sub-pixel power) may result in signal leakages between
different Stokes parameters.
The intensity signal is much higher than the polarization signal, and therefore 
any sub-pixel power which is a small fraction of the intensity may be substantial 
when compared with the polarization signal.  By over-pixelizing the intensity we 
can lower sub-pixel power and mediate the leakage of this power to the $Q$ and $U$ 
basis.  In this way we reduce the aliased sub-pixel intensity signal present in the 
estimated $Q$ and $U$ maps.  

Such an effect has been indeed observed in simulated cases~\citep{bib:ashdown07a}.
MADmap's ability to accommodate different pixel resolutions for different 
Stokes parameters allows us to robustly bypass such problems by reducing 
the size of the $I$ pixels and with it the amount of the power leaked from
$I$ to $Q$ and $U$.

We note that all these capabilities are unique compared to other CMB 
map-making software packages.

\section{MADmap Map-Making Examples} \label{sec:examples}

We will describe the tests that have been performed on MADmap and
highlight the functionality provided by the M3 data abstraction layer,
and the Generalized Compressed Pointing (GCP) library.  
MADmap has been run on an array of real and simulated data.  In this
paper, we would like to describe two large runs that demonstrate MADmap's
capabilities and capacity.  Both of these runs are analyses of
simulated data, the first a year-long Planck satellite 
mission~\citep{bib:planck05}, and the second a two-week EBEX balloon-borne
measurement \citep{bib:oxley04, bib:grainger08}.
In the following we start from a short description of features of each 
of the experiments and their data set and continue with a detailed discussion
of MADmap map-making run performance.

\subsection{Planck}
Planck is a ESA/NASA satellite, which will carry on board two instruments:
Low and High Frequency (LFI and HFI, respectively) and a total of $74$
independent detectors ($54$ of which are polarization-sensitive) measuring the sky signal in $9$ frequency bands.
Over a period of fourteen months Planck will observe the full sky twice\footnote{A decision on extending the mission to support two further coverings of the sky is pending.} at a 
resolution dependent on the frequency band, ranging from $33'$ at $30$GHz down to $5'$ at $217$GHz and above.
The detector sampling frequency likewise ranges from $32.5$Hz at $30$GHz up to $172$Hz at $100$GHz and above . 
Consequently, the Planck time ordered data set will consist of roughly 325 Giga-samples, from which $23$ maps will be derived (where a map is counted for each stokes parameter and each 
observing frequency). The number of sky pixels per map will be as high as 50 Mega-pixels for the high frequency, high resolution channels. With only two full sky coverings, and modest sampling rates, Planck will achieve a relatively low density of observations per unit sky area.  This results in a relatively low signal-to-noise ratio at the resolution scale
of the recovered maps, particularly when considering the faint magnitude of the polarization signal.

The Planck simulations used here were calculated with the Planck Level S 
simulation pipeline \citep{bib:reinecke06}.  By default this software generates and writes to disk telescope pointing files and
detector data time streams. For some of the tests, we used the Level S package 
to simulate and store only the telescope pointing information, while the 
time domain data themselves were generated only at run-time using the M3 on-the-fly
simulation capability. The full Planck focal plane analysis was the largest scale MADmap run 
to date, which turned $325$ Giga-samples into estimates of $150$ Mega-pixel amplitudes.

\subsection{EBEX}

EBEX is a balloon-borne experiment which will collect data during a roughly two-week long circumpolar flight in Antarctica.
It will carry on board 1406 detectors observing in 3 different frequency
bands: $140$, $220$ and $410$GHz. The resolution at each frequency will 
be $8'$.
The sampling frequency is set to $190$Hz. During its flight, EBEX 
will target a small area amounting to roughly $1\%$ of the entire sky. 
Due to its much larger number of detectors and higher sampling rate, the total length of the EBEX time ordered data set is comparable to the Planck baseline mission. 
The adopted observation strategy will result in a very deep integration 
of the probed sky area reaching up to millions of measurements per beam-size 
pixel. Consequently, the recovered maps 
will have high signal to noise ratio at the resolution scale. 
This is driven by the scientific goals of EBEX, which are to detect and 
characterize the minuscule CMB B-mode polarization signature.
The size of the maps produced from the EBEX data will be limited
to $10^5$ pixels, i.e. half a percent of one of the Planck maps. 
For the EBEX simulations we use the M3
simulation capability to generate time stream data on the fly when
MADmap makes data requests.

\subsection{Performance analysis}
The simulations presented here were conducted at the Department of
Energy NERSC supercomputing center on the Cray XT4 named Franklin.
Franklin has 9660 compute nodes, and at the time when the Planck
full focal plane simulation was run in September 2007 these nodes contained one 2.6 GHz
clock speed dual core AMD Opteron processor with 2 gigabytes of
memory per core.  When the scaling tests for the Planck and EBEX simulations were run in October 2008 the
Franklin processors had been updated to quad core AMD Opteron
processors running at 2.3 GHz with two gigabytes of memory per core.
Franklin has a fast SeaStar2 switch interconnect, and runs the Lustre
file system to control 350 terabytes of usable disk space.

We performed a sequence of scaling tests on MADmap to show its
performance on large data sets and computing at scale.  These runs are
broken into three cases.  The first two are nearly identical
simulations of the Planck telescope's 217 GHz channel.  For one of
these simulations the data are precomputed and read from disk, and in
the other case the simulation of the time stream is done at run time.
The third case is a simulation of the EBEX experiment with data simulated at run time.  Through these three simulations we show the
performance on high resolution full sky data in the Planck case, and a
small patch medium resolution very low pixel noise simulation in the 
EBEX case.  
By simulating data at the time of request we are demonstrating MADmap's
capacity for Monte Carlo simulation analysis with negligible disk use.
We also present an analysis of a year of data collected by the
entire Planck focal plane as a demonstration of computational capacity
in both the pixel and time domains.

In the scaling runs each of the three
cases was analyzed at four different processor concurrencies.  For the
Planck analysis these concurrencies were 2,196, 4,392, 8,784 and 17,568
processor cores, and the EBEX analyses were run on 1,920, 3,840, 7,680,
and 15,360 processor cores.  In the Planck scaling runs the number of time samples,
${\rm n_t}$, was  $7.6\times 10^{10}$ and the number of observed pixels in all three polarization
maps, ${\rm n_z}$, was $1.5\times 10^8$.  In the EBEX case ${\rm n_t}$ was $1.6\times 10^{11}$ and ${\rm n_z}$ was $1.0\times 10^5$.

\begin{figure*}[htbp]
\begin{center}
\includegraphics[scale=.5]{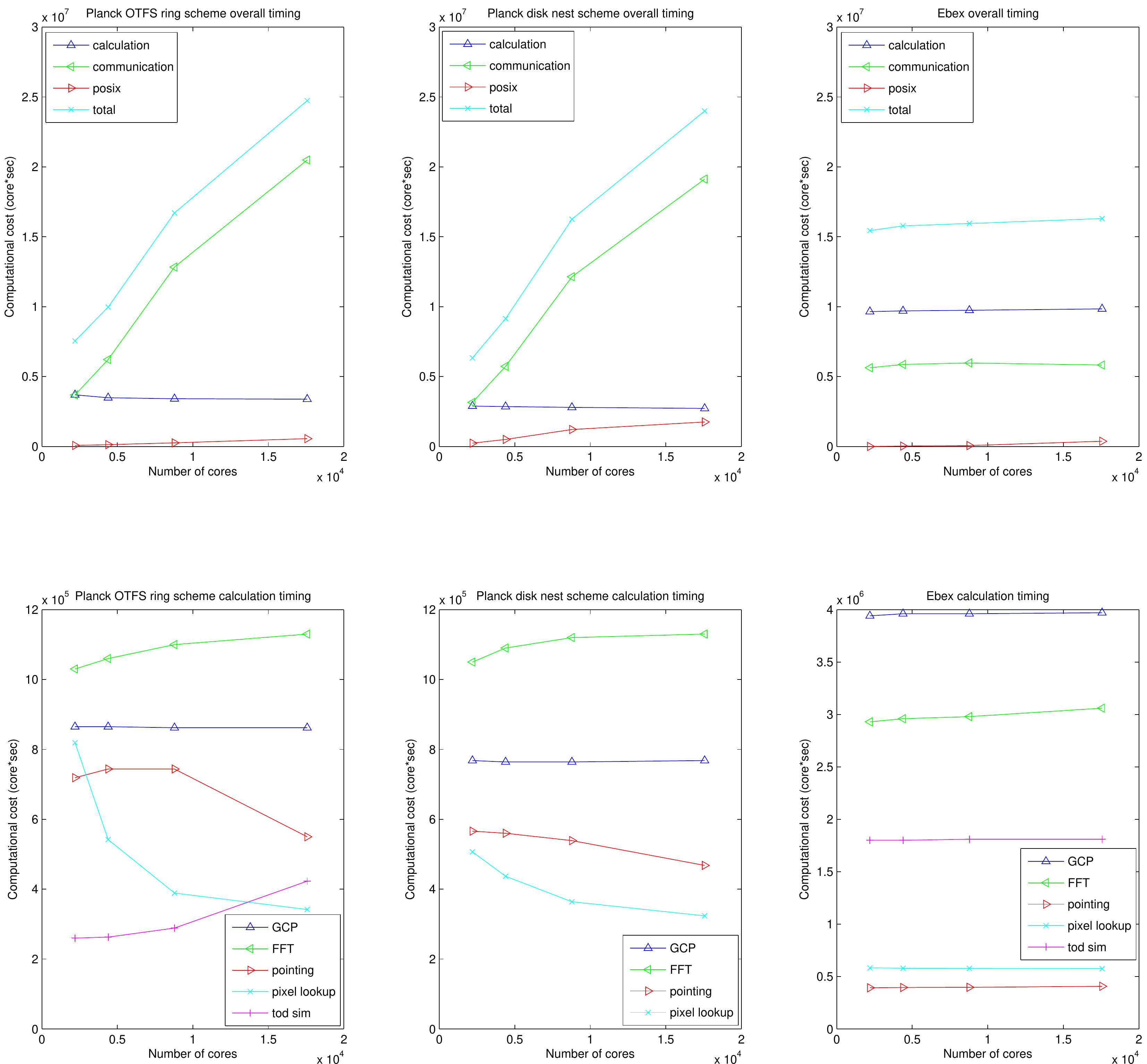}
\caption{This figure plots the results of the three scaling tests that
  were performed.  Each run is represented by two plots.  One measures
  the overall cost of calculation, communication and I/O; the other
  breaks down the time spent doing computations between the algorithms
  that comprise the computations.
  \label{labelScalingPlot}
}
\end{center}
\end{figure*}

The results of these runs are shown in Figure~\ref{labelScalingPlot}.  There are some particular features that
stand out in the plots.  These are strong scaling plots, therefore measured in
processor seconds, and ideally these would be flat, implying total cost
invariance with processor concurrencies.  The cost of the computation
is close to this ideal, which implies that the problem is well
balanced.  The input/output subsystem is very limited under the
strain of large concurrencies of file requests and does not scale
well.  This is a known problem with contemporary supercomputers run at
scale, and highlights the power of being able to simulate data at the time
of request in order to avoid the I/O bottle neck.  This point is clear
when the time spent in I/O in the Planck disk case is
compared to that of the Planck on-the-fly simulation case.

Communication cost for all sky high resolution maps is significant,
but in the low pixel count case of EBEX the communication is
sub-dominant.  The communication cost is essentially linear with the
number of observed pixels.  The EBEX case has just over 100 thousand
pixels observed, and the Planck case has 150 million pixels observed.
Over the concurrencies probed, the range of pixels spans the space
where communication is significantly smaller and larger than the
computation cost.  The Planck case scaled up to over 10k processors
shows the limitations of the parallelism used for the pixel domain
data.  Work is ongoing to improve this limitation so that MADmap will
scale to hundreds of thousands of processors and hundreds of millions
of pixels.

The computation of the pixel look-up actually requires fewer
calculations at higher concurrencies because the complexity is a
function of the number of locally observed pixels which decreases as
the time stream is divided more finely.  We can also see the effects
of memory locality and cache misses in the computation of the
pointing.  This is highlighted by comparing the Planck runs done with
the Healpix nested scheme and the Healpix ring scheme \citep{bib:gorski05}.  
In the nested scheme, pixels nearby
on the sky have similar pixel indexes, and therefore show better cache
performance in the calculation of the action of the pointing matrix.

\begin{figure*}[htbp]
\begin{center}
\includegraphics[scale=.5]{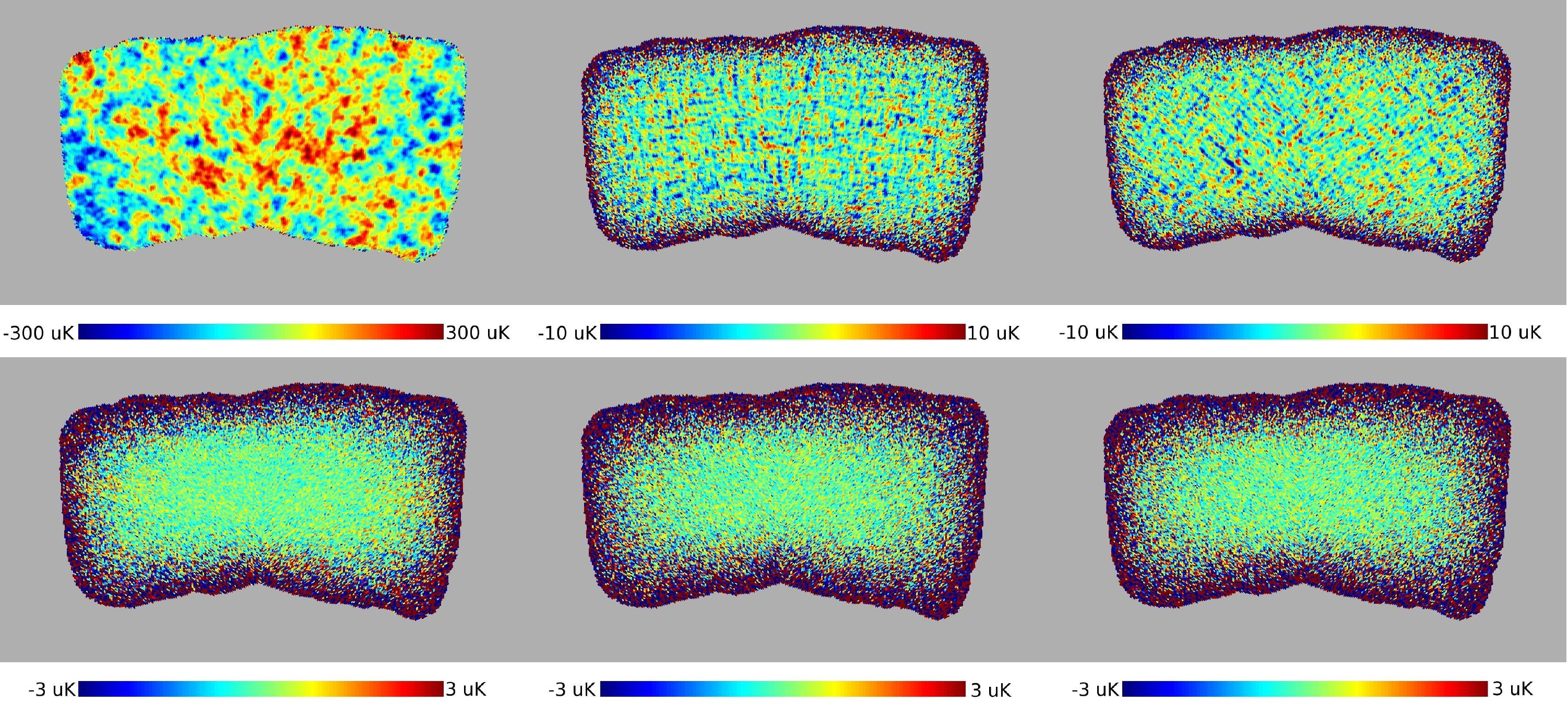}
\caption{This figure shows the map resulting from the EBEX scaling
  runs, which recovered the polarized sky signal from $1.6\times 10^{11}$
  time samples.
  The color scale is given in units of $\mu K$ thermodynamic
  temperature.  Plotted in the top three panels are the I, Q and U
  polarization components, and below each map is an image of the
  difference with the input signal map used to generate the
  simulation.  
\label{ebexMapIQU}
}
\end{center}
\end{figure*}

\begin{figure*}[htbp]
\begin{center}
\includegraphics[scale=.5]{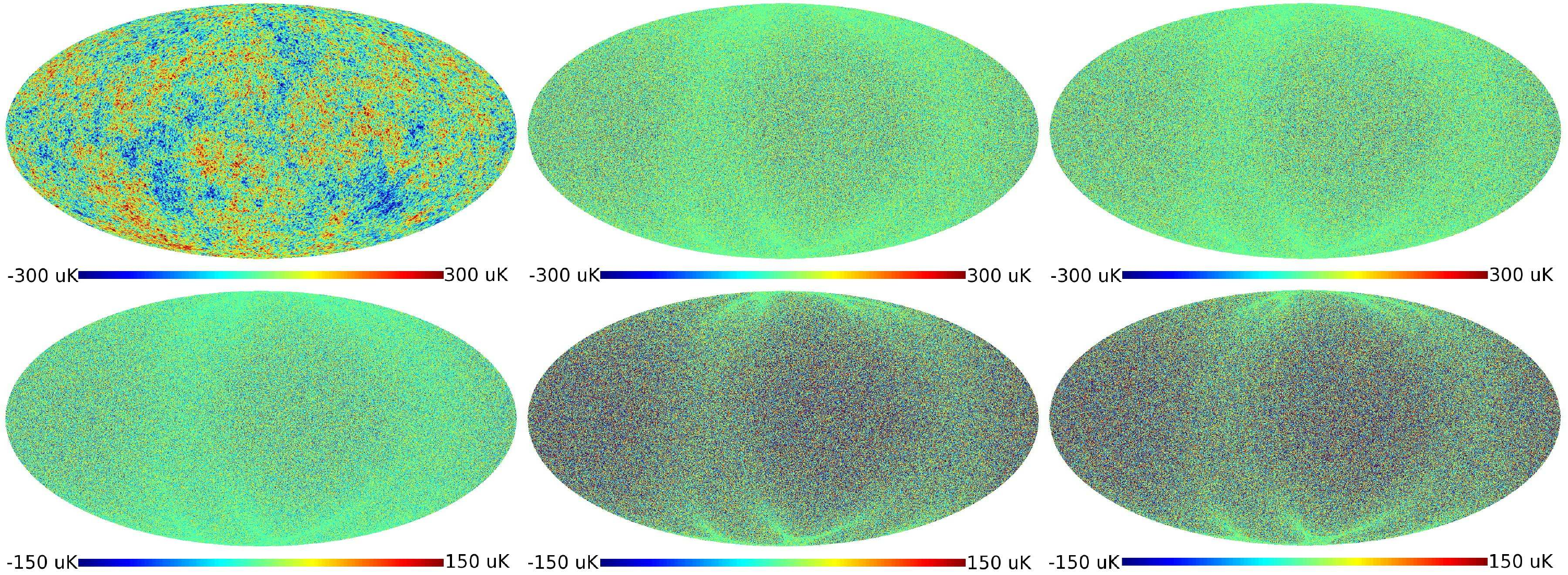}
\caption{This figure shows the map resulting from the Planck scaling
  runs.  The color scale is given in units of $\mu K$ thermodynamic
  temperature.  Plotted in the top three panels are the I, Q and U
  polarization components, and below each map is an image of the
  difference with the input signal map used to generate the
  simulation.  Note that for the Planck 217 GHz channel the Q and U
  maps are noise dominated when produced at high resolution (Healpix
  nside parameter of 2048). The data set simulated for this run comprised
  of $7.5\times10^{11}$ samples divided between $74$ detectors and $9$ 
  frequency  bands.
  \label{planckMapIQU}
}
\end{center}
\end{figure*}

In the case of the Planck full focal plane run described earlier, 
which constituted the largest MADmap run  performed to date,
the analysis was done in the
concatenated data distribution in low memory mode on 16,384 cores and
completed in 32 minutes.  In this run, the expansion of
the compressed pointing by the GCP library consumed 19.0\% of the
run time.  The calculation of FFT's with the ACML library consumed
11.9\% of the wall clock time.  Pixel re-indexing constituted 6.3\% of
run time and multiplying by the sparse pointing matrix 
4.9\%.  Outside of these calculations, 24.0\% of run time was spent
on communication of signal vectors, 23.1\% on reading simulated
time streams from disk, and 3.8\% on writing maps to disk.  

A sample of the results of the MADmap runs analyzing the simulated 
EBEX and Planck data sets is presented in 
Figures~\ref{ebexMapIQU}~\&~\ref{planckMapIQU}, respectively.

\section{Comparison With Other Codes} \label{sect:comparison}

Map-making is one of the principal constituents of the data analysis pipeline
of any CMB experiment. Consequently, there has been a significant
body of work devoted to different algorithmic and implementation aspects
of the map-making problem, 
~\citep[e.g.,][]{bib:wright96,bib:janssen96,bib:tegmark97,bib:delabrouille98,  bib:borrill99, bib:oh99,bib:maino99,bib:dore01,bib:stompor02,bib:hinshaw03,bib:keihanen04,
bib:stompor04,bib:armitage04,bib:degasperis05,bib:keihanen05,bib:poutanen06,bib:patanchon08,
bib:armitage08,bib:sutton09,bib:kurki-suonio09}, and many numerical map-making codes have been 
developed over the years and reported in the literature. Few of those have ever been 
implemented for a massively parallel architecture. Still fewer of those have been applied to the data sets of the size
and complexity as expected from Planck and the next generation of the CMB 
experiments. Only some of those codes are flexible and general enough to be 
applicable to a realistic CMB data set as produced by scanning experiments. To the best 
of our knowledge none allows for fully arbitrary pointing matrices as MADmap does,
instead usually relying on its hard coded, parametrized version appropriate for some
specific instrument. Moreover, although some of the existing codes are used 
by larger groups of researchers, none is readily available in the public domain.

Over the past few years the Planck CTP working group has undertaken a coordinated effort 
to compare some of the existing map-making codes. The final stage the comparison 
involved 5 codes, including 2 Planck-specific codes referred to as 'destripers' and 3 
maximum-likelihood codes including MADmap. The aim of these comparisons was to first 
demonstrate the software's capacity to deal with the requisite volume of data, and then the precision
of the results, and the resource requirements were measured and compared.  

The results have been published in series of papers~\citep{bib:ashdown07a,bib:ashdown07b,
bib:ashdown09}. Those results validated MADmap, which has been shown to produce
maps virtually indistinguishable from those of the other maximum-likelihood 
map-makers and the produced maps were consistent with input maps used to generate the time ordered data simulations.
The MADmap results were also found to be very similar to those of the destripers, though MADmap was somewhat closer to the input map.  
This small advantage of the maximum-likelihood codes
is understandable and expected as they are designed to produce minimum variance, optimal
maps. In terms of the resources, the superior CPU and memory performance of
MADmap with respect to other maximum-likelihood estimators has been
duly demonstrated in~\citet{bib:ashdown09}. The destripers, devised as Planck-specific 
map-makers \citep[but see][]{bib:sutton09}, gain on speed and/or memory requirements 
at the price of their generality, and therefore were capable of delivering a comparable 
performance to MADmap at the fraction of its resource requirements. We note that the resource 
comparison presented in~\citet{bib:ashdown09} was obtained using a data set of a rather 
modest size with a total number of samples ${\rm n_t}\sim 10^9$ and pixels ${\rm n_z}\sim9\times10^6$ 
including maps of the three polarization Stokes parameters.  

In the particular simulation presented in~\citep{bib:ashdown07b}, when run in large memory mode, MADmap was found to 
require less than half of the CPU resources of the other two maximum likelihood 
map-makers in the comparison (44\% of the CPU resources required by MapCumba, and 40\% 
of the CPU resources required by ROMA), while MADmap had a slightly higher memory 
footprint.  MADmap's slightly higher memory usage is because MADmap stored a different noise filter for each stationary period where
the other codes assumed a single noise filter for the entire data set.  
When run in low memory mode MADmap's memory footprint was nearly one quarter the
size of the other two maximum likelihood implementations, and the CPU resources required were
about two thirds of those required by the other maximum-likelihood solvers (68\% of the CPU 
resources required by MapCumba, and 61\% of the CPU resources required by ROMA).  
This paper complements the results published in the CTP papers by showing
the scalability of the MADmap code to the large concurrences and its performance
while facing the most voluminous forthcoming data sets as well as highlighting its
additional features not tested in the CTP runs.

The majority of the codes and methods discussed above work efficiently, or typically only,
in pixel domain. Harmonic space map-makers have been also recently 
considered~\citep{bib:armitage04,bib:armitage08} though demonstrated so far only in the
cases with uncorrelated time-domain noise. The potential advantage of the harmonic
approach is that it allows to correct for a potential beam asymmetry in a more straightforward
manner. We note that such a beam correction can be at least to some extent resolved
also in the pixel domain~\citep{bib:burigana03} though would require an appropriately defined pointing
matrix. Given that, MADmap can be a tool of choice for performing such a study in the
future.

\section{Conclusions  \& Future Work} \label{sec:conclusions}

This paper offers an in-depth description of the MADmap software
application, describing the details of the algorithm and the massively
parallel implementation.  
What is presented in this paper is MADmap's
capacity to scale to very large problems and very large processor
concurrencies.  We discussed the variety of modes in which MADmap can
be run, which allows for trade-offs between CPU and memory
consumption.  One of the main features that sets MADmap apart from
other map-makers is the wide flexibility in the data model that
enables support for any sparse linear model of the signal to be
estimated.  This highly adaptable model for the signal is further
enabled by the GCP library which provides a modular plug in
architecture that can be used to compute the sparse linear model from
compressed data.  The functionality provided by GCP is what enables
MADmap's low memory mode.

Recall that, in low memory mode, the sparse pointing matrix
is created in a buffered way from cached compressed data through the
GCP library.  When MADmap is run in the stacked data distribution, the
compressed data can be reused for multiple detectors, and in this way
reduces memory and sometimes CPU requirements for GCP.  GCP's
modularity allows for the addition of features which can be used to
accomplish a wealth of science goals.  One interesting use is the
calibration with a known signal (like the CMB dipole) as a byproduct
of map-making while consistently accounting for the extra degree of
freedom added by this estimation.  Another possibility is to do
component separation during map-making to produce maps of different
astrophysical processes from time streams of data collected at
different observing frequency bands.  This can be accomplished if the
linear scaling terms for the projection from frequency band to
astrophysical component are known \textit{a priori}, and a GCP module
is added to produce a sparse pointing matrix that corresponds to this
projection.  A capability which currently exists is the removal of
systematic effects corresponding to a linear scaling of a measured
template.  These systematic effects range from the temperature of the
cold stage of the cryogenic electronics, to a measured template for
thermal pick-up from the ground as a function of a ground-based
telescope's orientation, or any other \textit{a priori} known signal
to be marginalized from the output maps.  GCP also promises the
ability to deconvolve compact beam effects during map-making by
representing some portion of the beam profile in the pixel domain in
the sparse pointing matrix.  This can be used to account for
asymmetries in the beam, through the combining of detectors with
disparate beam profiles into a single map (usually required for
component separation during map-making), or can even be used to
deconvolve in the pixel domain the effects of a bolometric detector's
response curve in the time domain.  

MADmap is capable of solving massive problems quickly and efficiently
given appropriate computational resources and such large runs have
been presented in this paper.  MADmap is a useful and efficient tool
that can be run on much more modest computing resources, and the scale
of the runs presented should not misinform the reader about minimum
resource requirements of MADmap.  MADmap can be quite computationally
efficient on small cluster sized resources, and when run in low memory
mode it can reduce large time ordered data sets in core memory.  The
limiting factor in this case is run time, but at low concurrencies,
and when using a reasonable communication fabric, MADmap is very
efficient.  In addition, often on these small clusters a limited user
pool allows for the execution of jobs with very long wall clock times,
and in these situations MADmap can tackle non-trivial problems with a
modest computer.  MADmap implements efficient check-pointing, and this
allows for the exploitation of long-running jobs without fear of
wasting resources in the case of a system failure during execution.

The massively parallel computing community is on a road map to push its systems
to the exascale, which is $10^{18}$ calculations per second, by about 2018.  
The primary impediment to scaling up MADmap to the exascale and beyond
is the collective communication calls used to reduce maps.  It is the
subject of ongoing research as to optimize our collective communication
to take best advantage of character of our data distribution, its sparsity and the communication 
fabric topology.  

MADmap does not currently have a facility for modeling correlations in
the noise that may exist between different data sets, which are to be processed 
simultaneously, such as measurements coming from different detectors of the
same experiment.  Such correlations may be due to electronic cross talk, thermal drifts of a
cryogenic focal plane, or in the case of ground based experiments,
atmospheric noise that is seen by the focal plane simultaneously.
This correlations can be very important to some collected data sets,
and incorporating this functionality is a focus of ongoing work.

In future publications we will present more applications of MADmap in 
particular those demonstrating the GCP library functionality on real and simulated 
data. A package 
containing MADmap and all its dependencies is available for download at 
the LBNL CodeForge website\footnote{LBNL CodeForge:\\ {\tt https://codeforge.lbl.gov/projects/cmb/files}}.
MADmap, GCP and M3 are free software and distributed under the GNU
public licence agreement with hope that it will become a software package of 
the choice for the CMB community for an efficient computation of the microwave 
sky maps.

\acknowledgments We are grateful to NASA (through the Planck Project) and DoE for funding and
facilities.  This research used resources of the National Energy
Research Scientific Computing Center, which is supported by the Office
of Science of the U.S. Department of Energy under Contract
No. DE-AC02-05CH11231.  Thanks to Charles Lawrence for his help 
and guidance in this project and the submission of this paper.  
We acknowledge the use of simulations of
telescope pointing provided by the Planck and EBEX collaborations.  In
particular the simulation of the Planck mission was done by the
simmission \citep{bib:reinecke06} application primarily written by
Daniel Mortlock and managed by Martin Reinecke.  The sky scanned in
the Planck simulation was provided by the Planck collaboration and is
the model derived by the Planck foregrounds working group.  The EBEX
telescope pointing simulation was provided by Samuel Leach, and the sky
simulation was provided by Nicolas Ponthieu.  The parametrization of
the noise properties of EBEX was provided by Shaul Hanany, and the
simulation was generally a product of the EBEX data analysis working
group.



\appendix

\section{Proof that critical FFT length minimizes complexity of convolution algorithm}
\label{Appendix:optimalFFT}.
\begin{proof}  
In Section~(\ref{sec:weighting}) we need to show that $h''({\rm n_f}) > 0$ for our critical value of ${\rm n_f}$
so as to insure that our critical point minimizes the complexity.  Recall that $h''({\rm n_f})$ is 
defined as 
\begin{equation*}
h''({\rm n_f}) = \frac{2 {\rm n_f} \ln({\rm n_f})}{({\rm n_f}-2{\rm n_c})^3}-\frac{2(1+\ln({\rm n_f}))}{({\rm n_f}-2{\rm n_c})^2}+\frac{1}{{\rm n_f}({\rm n_f}-2{\rm n_c})}
\end{equation*}
Let us assume that $h''({\rm n_f}) \le 0$ and derive a contradiction 
\begin{eqnarray*}
\frac{2 {\rm n_f} \ln({\rm n_f})}{({\rm n_f}-2{\rm n_c})^3}-\frac{2(1+\ln({\rm n_f}))}{({\rm n_f}-2{\rm n_c})^2}+\frac{1}{{\rm n_f}({\rm n_f}-2{\rm n_c})} \le 0\\
\frac{1}{2{\rm n_f}} + \frac{{\rm n_f} \ln({\rm n_f})}{({\rm n_f}-2{\rm n_c})^2} \le \frac{1+ln({\rm n_f})}{{\rm n_f}-2{\rm n_c}}\\
({\rm n_f}-2{\rm n_c})^2 \le 2{\rm n_f}({\rm n_f}-2{\rm n_c})(1+\ln({\rm n_f}))-2{\rm n_f}^2\ln({\rm n_f})\\ 
4{\rm n_c}^2+4{\rm n_c}{\rm n_f}\ln({\rm n_f}) \le {\rm n_f}^2 
\end{eqnarray*}
substitute ${\rm n_c}$ in terms of the critical ${\rm n_f}$
\begin{eqnarray*}
{\rm n_c} = \frac{{\rm n_f}}{2(1+\ln({\rm n_f}))}\\
4\left(\frac{{\rm n_f}}{2(1+\ln({\rm n_f}))}\right)^2 + 4\left(\frac{{\rm n_f}}{2(1+\ln({\rm n_f})}\right){\rm n_f}\ln({\rm n_f}) \le {\rm n_f}^2\\
\frac{1}{(1 + \ln({\rm n_f}))^2} + \frac{2 \ln({\rm n_f})}{1+\ln({\rm n_f})} \le 1\\
1 + 2(1 + \ln({\rm n_f}))\ln({\rm n_f}) \le (1 + \ln({\rm n_f}))^2\\
1 + 2 \ln({\rm n_f}) + 2 \ln({\rm n_f})^2 \le 1 + 2 \ln({\rm n_f}) + \ln({\rm n_f})^2\\ 
\ln({\rm n_f})^2 \le 0\\
{\rm n_f} \le 1
\end{eqnarray*}
The length of the FFT must be at least 3 if ${\rm n_c}$ is non-zero, and we have derived a contradiction, therefore,
$h'({\rm n_f}) = 0 \Rightarrow h''({\rm n_f}) > 0$.
\end{proof}

\section{Table of variables used in text}
\begin{table*}[hbtp]
  \centering
  \begin{tabular}{|c|l|}
  \hline
  Symbol & Description \\
  \hline
  ${\rm c_h}$ & The aggregate number of operations required for high memory mode\\
  ${\rm c_l}$ & The aggregate number of operations required for low memory mode\\
  ${\rm c_e}$ & The aggregate number of operations required for extremely low memory mode\\
  ${\rm m_h}$ & The aggregate distributed memory requirement required for high memory mode\\
  ${\rm m_l}$ & The aggregate distributed memory requirement required for low memory mode\\
  ${\rm m_e}$ & The aggregate distributed memory requirement required for extremely low memory mode\\
  ${\rm m_{ns}}$ & The per processor memory requirement for the noise filters in the stacked distribution mode\\
  ${\rm m_{nc}}$ & The per processor memory requirement for the noise filters in the concatenated distribution mode\\
  ${\rm d_t}$ & The per processor disk requirement for reading time stream of data\\
  ${\rm d_{ps}}$ & The per processor disk requirement for reading compressed pointing in the stacked distribution mode\\
  ${\rm d_{pc}}$ & The per processor disk requirement for reading compressed pointing in the concatenated distribution mode\\
  ${\rm d_{pd}}$ & The per processor disk requirement for reading detector pointing without compressed pointing\\
  ${\rm d_{ns}}$ & The per processor disk requirement for reading noise filters in the stacked distribution\\
  ${\rm d_{nc}}$ & The per processor disk requirement for reading noise filters in the concatenated distribution\\
  ${\rm n_z}$ & The number of pixels in the map\\
  ${\rm n_d}$ & The number of detectors.   \\
  ${\rm r_d}$ & The detector sample rate\\
  $\Delta t$ & The integral time over which the telescope observed data\\
  ${\rm n_t}$ & The total number of samples measured by all detectors. \\
  ${\rm n_i}$ & The number of iterations required to solve the PCG \\
  ${\rm n_c}$ & The correlation length of the noise\\
  ${\rm n_b}$ & The number of stationary periods\\
  ${\rm r_p}$ & The telescope telemetry data sample rate\\
  ${\rm n_{nz}}$ & The number of non-zero elements per row of the pointing matrix. \\
  ${\rm n_{proc}}$ & The number of processors used in a run. \\
  \hline
  \end{tabular}
  \caption{Some notes on the variables: The number of pixels in the
    map, ${\rm n_z}$, is more generally the number of free parameters in the
    signal model which could include signals in addition to the pixels
    in the map and may include multiple maps.  Note that for detectors
    sampled simultaneously and at the same rate ${\rm n_t} = {\rm n_d r_d} \Delta
    t$.  The number of iterations required to solve the PCG, ${\rm n_i}$ is
    proportional to the condition number of the pixel-pixel noise
    correlation matrix after preconditioning.  The correlation length
    of the noise, ${\rm n_c}$, is also the band width of the diagonal
    blocks of the inverse time-time noise correlation matrix.  The
    number of stationary periods, ${\rm n_b}$, corresponds to the number of
    diagonal blocks in the inverse time-time noise correlation matrix.
    The telescope telemetry data sample rate, ${\rm r_p}$, is more generally
    the minimum sample rate required to accurately reconstruct the
    pointing matrix $A$.  }
    \label{tab:param}
\end{table*}


\bibliographystyle{apj}
\bibliography{madmap}



%% \begin{thebibliography}
%% \end{thebibliography}

\end{document}